\documentclass[iop]{emulateapj}
\usepackage[usenames,dvipsnames]{color}
\usepackage{amsmath,amsfonts,latexsym,graphicx,amssymb,times,appendix,bm,multirow,dcolumn,xspace, aas_macros,verbatim}
\usepackage[normalem]{ulem}

\usepackage[breaklinks,colorlinks,citecolor=blue]{hyperref}
\newcommand{\SN}{\textsc{SkyNet}}
\newcommand{\Sw}{\textit{Swift}}

\newcommand{\nes}{\ensuremath{\mathrm{n\_estimators}}}
\newcommand{\mss}{\ensuremath{\mathrm{min\_samples\_split}}}
\newcommand{\mf}{\ensuremath{\mathrm{max\_features}}}
\newcommand{\lr}{\ensuremath{\mathrm{learning\_rate}}}

\newcommand{\rfrac}[2]{\ensuremath{{}^{#1}\!/_{#2}}}

\begin{document}

%\title{Modeling the \Sw{} BAT Trigger Algorithm with Machine Learning}
\title{Machine Learning Model of the Swift/BAT Trigger Algorithm for Long GRB Population Studies}

\author{Philip B. Graff}
\affiliation{Department of Physics and Joint Space Science Institute, University of Maryland, College Park, MD 20742, USA}
\affiliation{NASA Goddard Space Flight Center, 8800 Greenbelt Rd., Greenbelt, MD 20771, USA}
\email{pgraff@umd.edu}
\author{Amy Y. Lien}
\affiliation{NASA Goddard Space Flight Center, 8800 Greenbelt Rd., Greenbelt, MD 20771, USA}
\email{amy.y.lien@nasa.gov}
\author{John G. Baker}
\affiliation{NASA Goddard Space Flight Center, 8800 Greenbelt Rd., Greenbelt, MD 20771, USA}
\email{john.g.baker@nasa.gov}
\author{Takanori Sakamoto}
\affiliation{Department of Physics and Mathematics, College of Science and Engineering, Aoyama Gakuin University, 5-10-1 Fuchinobe, Chuo-ku, Sagamihara-shi, Kanagawa 252-5258, Japan}

\begin{abstract}
To draw inferences about gamma-ray burst (GRB) source populations based on \Sw{} observations, it is essential to understand the detection efficiency of the \Sw{} burst alert telescope (BAT). This study considers the problem of modeling the \Sw{}/BAT triggering algorithm for long GRBs, a computationally expensive procedure, and models it using machine learning algorithms. A large sample of simulated GRBs from~\cite{Lien2014} is used to train various models: random forests, boosted decision trees (with AdaBoost), support vector machines, and artificial neural networks. The best models have accuracies of $\gtrsim97\%$ ($\lesssim 3\%$ error), which is a significant improvement on a cut in GRB flux which has an accuracy of $89.6\%$ ($10.4\%$ error). These models are then used to measure the detection efficiency of \Sw{} as a function of redshift $z$, which is used to perform Bayesian parameter estimation on the GRB rate distribution. We find a local GRB rate density of $n_0 \sim 0.48^{+0.41}_{-0.23} \ {\rm Gpc}^{-3} {\rm yr}^{-1}$ with power-law indices of $n_1 \sim 1.7^{+0.6}_{-0.5}$ and $n_2 \sim -5.9^{+5.7}_{-0.1}$ for GRBs above and below a break point of $z_1 \sim 6.8^{+2.8}_{-3.2}$. This methodology is able to improve upon earlier studies by more accurately modeling \Sw{} detection and using this for fully Bayesian model fitting. The code used in this is analysis is publicly available online\footnote{\url{https://github.com/PBGraff/SwiftGRB_PEanalysis}}.
\end{abstract}

%\pacs{}

\maketitle

%\tableofcontents

%++++++++++++++++++++++++++++++++++++++++++++++++++++++++++++++++++++++
%
%  INTRODUCTION
%
%++++++++++++++++++++++++++++++++++++++++++++++++++++++++++++++++++++++

\section{Introduction}\label{sec:intro}
%\note{Introduce the problem -- in context of both \Sw{} and machine learning.}

Long gamma-ray bursts (GRBs) are related to core-collapse supernovae from the death of massive stars. These are important for studying star-formation history, particularly in the early universe where other methods become difficult. The \Sw{} space telescope~\citep{Gehrels2004} is able to detect and localize these out to large distances and quickly downlink the data to the ground. These abilities enable prompt ground-based followup observations that can provide redshift measurements of the GRBs. To date, \Sw{} has detected over 900 GRBs, of which $\sim 30\%$ have redshift measurements. From these observations, one can try to infer the intrinsic GRB rate that is connected to stellar evolution over the history of the Universe. Many researchers have used \Sw{}'s observations to study intrinsic GRB redshift and luminosity distributions, and the implications for star-formation history \citep[e.g.,][]{Guetta07_Ibc,Guetta07_sfr,Yuksel08,Kistler08_SFR,Butler10,Robertson12,Pelangeon08,Salvaterra09,Campisi10, Wanderman10,Virgili11, Qin11,Salvaterra12,Coward13,Kanaan13, Wang13, Lien2014,Howell2014,Yu15, Petrosian15, Pescalli15}.

Several studies have suggested that the GRB rate at high redshift ($z \gtrsim 5$) is larger than the expectation based on star-formation rate (SFR) measurements \citep[e.g.,][]{Le07,  Yuksel08, Kistler09,Butler10,Ishida11,Tanvir12,Jakobsson12,Lien2014}. This result could imply several possibilites, such as a larger star-formation rate in the early universe\citep[e.g.,][]{Kistler09, Tanvir12}, an evolving luminosity function\citep[e.g.,][]{Virgili11,Pescalli15}, or a different GRB to supernova ratio (i.e. a different scenario of stellar evolution) due to a different environment in the early universe \citep[e.g.,][]{Woosley12}.

However, it remains difficult to constrain the GRB rate. Though \Sw{} has observed a large population of GRBs only some of these have measured redshifts. Even with a relatively complete redshift sub-sample, there are complicated selection effects from the complex trigger algorithm adopted by the burst alert telescope (BAT) on-board \Sw{} and the difficulty in searching through a large parameter space. It is challenging to distinguish the luminosity function and the redshift distribution using the observational data. We address some of these issues with a machine learning approach to produce a fast, but reliable, treatment of \Sw{}'s instrumental selection effects, thereby enabling a robust Bayesian treatment of population model analysis.

%However, it is difficult to infer the intrinsic GRB rate as the \Sw{} trigger algorithm consists of over $500$ criteria based on photon count and an image threshold. A program capable of simulating this complex trigger algorithm was created in a previous work~\cite{Lien2014}, but the computational cost of simulating even a single GRB -- approximately 1 CPU min -- prohibited a larger study.

Machine learning (ML) is a field of research that involves designing algorithms (MLAs) for building models that learn from generic data. The models are fit to a set of training data in order to make predictions or decisions. Often, the original training data come from actual observations or simulations of a complex process. The models trained by MLAs can be evaluated very quickly for any new example after a one-time cost of training the model.

In this study, we look to aid the analysis of GRB data by using MLAs to train models that emulate the \Sw{} trigger algorithm. Our training data comes from simulations of GRB populations computed by~\cite{Lien2014}.

The structure of this paper is as follows. In Section~\ref{sec:swift_pipe} we describe the aspects of \Sw{} and its model for triggering on incident GRBs that are relevant to GRB population inferences. Then, in Section~\ref{sec:ML} we describe the machine learning algorithms used and compared in this study. Section~\ref{sec:results} presents the results of training the different ML models on the training data from the \Sw{} pipeline. We apply a trained ML model for accelerating Bayesian inference with faster likelihoods in Section~\ref{sec:bayesinf}, fitting the parameters of the intrinsic GRB rate distribution. Section~\ref{sec:comparison} compares our study to previous work estimating the intrinsic distributions of long GRBs with \Sw{} observations. Lastly, in Sections~\ref{sec:conclusion} and~\ref{sec:future} we summarize and propose future projects to follow-up.

%++++++++++++++++++++++++++++++++++++++++++++++++++++++++++++++++++++++
%
%  SWIFT DETECTION ALGORITHM
%
%++++++++++++++++++++++++++++++++++++++++++++++++++++++++++++++++++++++

\section{The \Sw{} Detection Algorithm}\label{sec:swift_pipe}

The burst alert telescope (BAT) on-board Swift adopts over 500 rate trigger criteria based on photon count rate in the raw light curve. Moreover, the burst needs to pass the ``image threshold'' determined by the signal-to-noise ratio estimated from an image generated on-board based on the duration found by the rate trigger criteria. Each rate trigger criterion uses a different energy band, different part of the detector plane, and different foreground and background durations for calculating the signal-to-noise ratio. In addition to the rate trigger algorithm, the BAT also generates an image every $\gtrsim$ minute to search for bursts that are missed by the rate trigger method (which is the so-called ``image trigger'')~\citep{Barthelmy05, Fenimore03, Fenimore04, McLean04, Palmer04}.

This complex trigger algorithm successfully increases the number of GRB detections. However, it also increases the difficulty of estimating the detection theshold, which is curcial for probing many intrinsic GRB properties from the observations. To address this problem,~\citet{Lien2014} developed a code that simulates the BAT trigger algorithm, and used it to study the instrinsic GRB rate and luminosity function. This ``trigger simulator'' follows the same trigger algorithm and criteria for the rate trigger as those adopted by the BAT, and mimics the image threshold and image trigger (see~\citet{Lien2014} for detailed descriptions). Although the trigger simulator can be used to address the complex detection thresholds of the BAT, it takes $\sim 10$ seconds to a few minutes to simulate the trigger status of a burst using a common PC with the 2.7 GHz Intel Core processor (the speed mainly depends on the number of bins in the light curve). Therefore, it is computationally intensive to perform a large number of simulations to cover a wide parameter space. This is where machine learning is able to accelerate our analysis.

%++++++++++++++++++++++++++++++++++++++++++++++++++++++++++++++++++++++
%
%  MACHINE LEARNING ALGORITHMS
%
%++++++++++++++++++++++++++++++++++++++++++++++++++++++++++++++++++++++

\section{Machine Learning Algorithms}\label{sec:ML}
To generate a fast emulator for the \Sw{} trigger simulator, we consider a variety of supervised learning algorithms, where the goal is to infer a function from labeled training data. Each example consists of input properties which are used to predict the output label.

Here we briefly describe each of the machine learning algorithms used in this study. We denote the set of input features by $\bm{x}$ and the machine learning model's predicted output is given by $y(\bm{x})$. The inputs are a set of $15$ parameters describing the GRB and detector as detailed in Table~\ref{tab:GRBparams}. Depending on the MLA, the output may be a discrete label, e.g. $\{0,1\}$, or it may be a continuous probability in $[0,1]$ and is designed to be the probability that a GRB, as specified by the features in $\bm{x}$, is detected by \Sw{}'s BAT. The true output is given by $\bm{t}$ and is $0$ for a non-detection and $1$ for a detection.

\subsection{Random Forests and AdaBoost}\label{sec:DecisionTreeEnsembles}
\begin{figure*}[htbp]
\centering
\includegraphics[width=\textwidth,keepaspectratio]{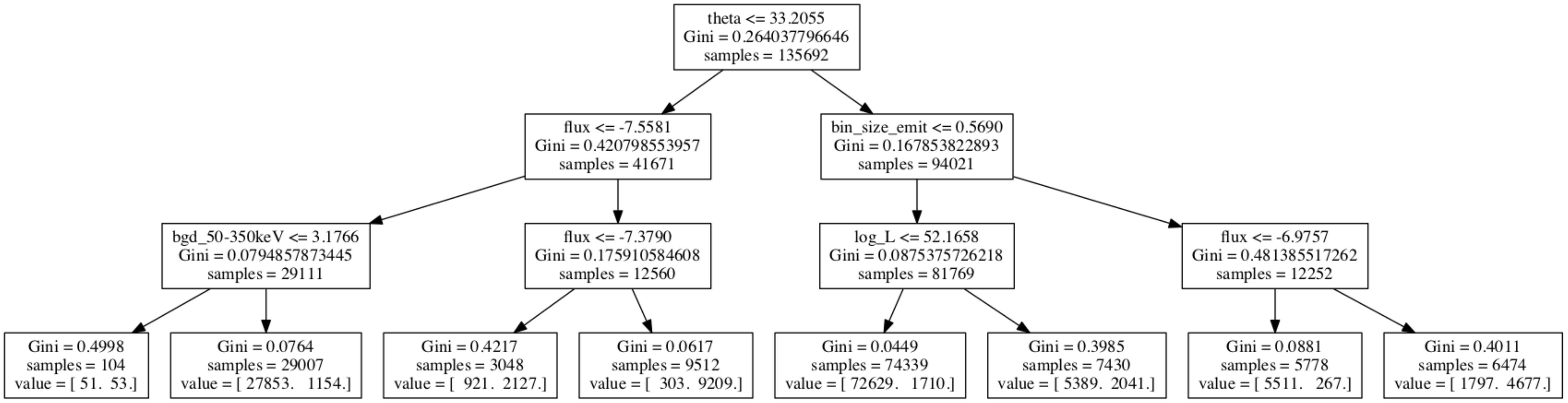}
\caption{A decision tree fit to the \Sw{} training data with a maximum depth of $3$. An accuracy of $93.9\%$ is achieved. Each box shows the parameter chosen for branching and the threshold used, as well as the total number of samples used to make that decision. The Gini factor shown is that from Equation~\eqref{eq:gini} for the subset at that location. At the leaves, the number of items with class 0 and class 1 are shown; the tree can assign class probabilities based on this split at the leaf that any new sample arrives at after following the branches down.}
\label{fig:decision_tree}
\end{figure*}

Random forests and AdaBoost both involve creating ensembles of decision trees, so we first introduce these as a machine learning model. In a decision tree, binary splits are performed on the training data input features, the dimensions of $\bm{x}$. In training a tree on data, a series of splits are made that choose a dimension and a threshold that optimize some criterion. Examples of this criterion are the accuracy of the resulting classifications (maximize; equivalently minimize errors) or the Gini impurity (minimize) which given by
\begin{equation}
G = 1 - \sum_{i=\{0,1\}} f_i^2,
\label{eq:gini}
\end{equation}
where $f_i$ is the fraction of correctly classified samples labeled with value $i$. This measure aims to make each sub-set resulting from a branch as ``pure'' as possible in the class labels of its members. Each split creates a pair of ``branches'', one with each class label. These splits are made until a stopping condition is reached (e.g. the samples are all of uniform class, a maximum number of splits has been reached, or the number of samples left to split among has fallen below a minimum value). This branch now becomes a ``leaf'' that assigns a class to all samples ending there. When a new event of unknown class is put into the tree, the tree will pass it through the learned splits/branches until it reaches a leaf, at which point it will be labeled according to the label of the leaf. An example tree fit to this data (with a hard limit of $3$ in depth) is shown in Figure~\ref{fig:decision_tree}; it has a classification accuracy of $93.9\%$ on the training data to which it was fit. Trees fit in the later models will be much larger and thus more accurate.

\subsubsection{Random Forests}\label{sec:RandomForests}
Random forests (RFs)~\citep{Breiman2001} improve upon classical decision trees by training an ensemble of trees that vote on the final classification. A ``strong learner'' (the RF) is created from an ensemble of ``weak learners'' (decision trees). In a RF, many decision trees are trained on the data -- often hundreds. To obtain many different trees, at each split in a tree, a random subset of the dimensions of $\bm{x}$ are chosen and the optimal binary split to be made out of these dimensions is made. Furthermore, each tree is trained on a bootstrap sample of the data; the original $K$ points are sampled with replacement to form a new set of $K$ points that may contain repeats. A RF thus guards against overfitting to the training data and potentially badly performing individual trees. A single tree can provide a probabilistic classification, $y_{\mathrm{DT}}(\bm{x}) \in [0,1]$, and combining many allows us to obtain a near-continuous probability, $y_{\mathrm{RF}}(\bm{x}) \in [0,1]$ by using
\begin{equation}
y_{\mathrm{RF}}(\bm{x}) = \frac{1}{N} \sum_{n=1}^{N} y_{\mathrm{DT},n}(\bm{x}),
\label{eq:RFprobfctn}
\end{equation}
with $N$ being the number of trees in the forest. This value we obtain as $y_{\mathrm{RF}}(\bm{x})$ is simply the probability that the GRB described by $\bm{x}$ is detected by \Sw{}.

%We use the implementation of RFs in the \texttt{randomForest} library~\cite{randomForestR} of the R statistical package. This allows us to use ($10$-fold) cross-validation to optimize for the number of randomly chosen features at each split. The default setting of $500$ trees in a forest is used as well as defaults for all other training parameters.
We use the implementation of RFs in the \texttt{scikit-learn}\footnote{\url{http://scikit-learn.org/stable/index.html}} Python library~\citep{scikit-learn}.

\subsubsection{AdaBoost}\label{sec:AdaBoost}
AdaBoost is short for ``Adaptive Boosting'', a meta-algorithm for machine learning~\citep{FreundSchapire1997}. It creates a single strong learner from an ensemble of weak learners, much like RFs. However, in the boosting framework, the decision trees are trained iteratively and when added together (as in Equation~\eqref{eq:RFprobfctn}) are weighted, typically based on their accuracy. Additionally, unlike for RFs, the training examples will not be all equally weighted when evaluating the accuracy. After an individual decision tree is added to the ensemble, the training data is reweighed so that examples that are misclassified increase in weighting and those classified correctly decrease in weighting. Therefore, future decision trees will attempt to better fit examples previously misclassified. In this way, the overall ensemble prediction may become more accurate.

Boosting may be applied to any machine learning algorithm, but in this work we apply it only to the decision tree weak learner (the other classifiers qualify as strong learners on their own and would thus likely not benefit significantly from boosting).
%We use the implementation of boosted classification trees in the \texttt{ada} library~\cite{adaR,Friedman2000} of the R statistical package. Cross-validation is used to optimize the number of trees trained and the maximum depth of the trees; the learning rate is kept at the default, $\nu=0.1$.
We use the implementation of AdaBoost for decision tree classifiers in \texttt{scikit-learn}. We note that the predicted probability, $y_{\mathrm{AB}}(\bm{x}) \in [0,1]$, is approximately continuous, similarly to $y_{\mathrm{RF}}(\bm{x})$.

\subsection{Support Vector Machines}\label{sec:SVMs}
Support vector machines (SVMs)~\citep{Cortes1995} are a tool for binary classification that finds the optimal hyper-plane for separating the two classes of training samples. Events are classified by which side of the hyper-plane they fall on. The hyper-plane that maximizes the separation from points in either class will (in general) have minimal generalization error for new data points.

\begin{figure}[htbp]
\centering
\includegraphics[width=0.7\columnwidth,keepaspectratio]{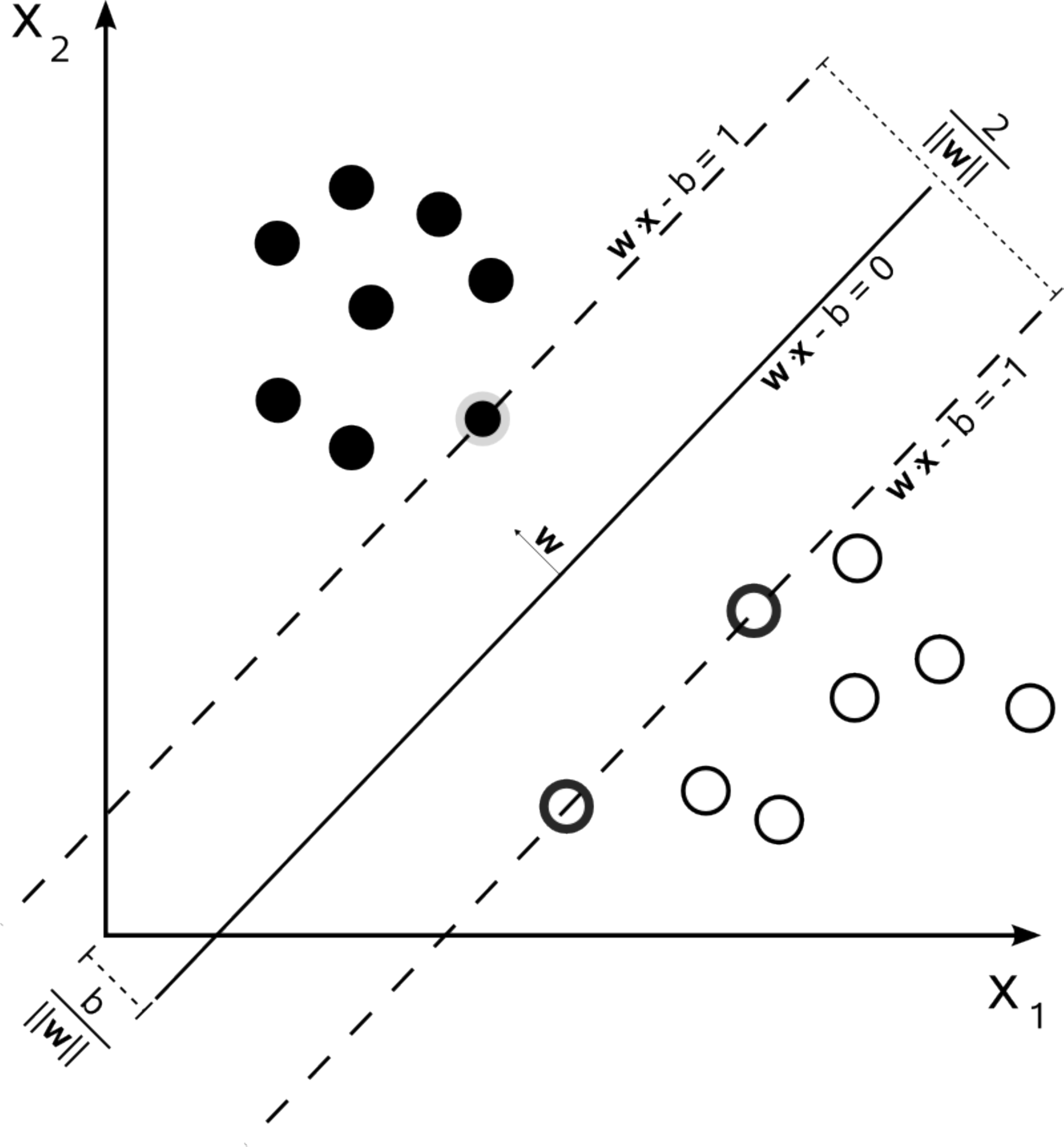}
\caption{The maximum separating hyper-plane and the margin hyper-planes for a toy data set. The ``support vectors'' are the highlighted points along the margin hyper-planes. Image courtesy of Wikimedia Commons~\citep{SVMtoyimage}.}
\label{fig:SVMtoy1}
\end{figure}

In a linear SVM, we label the two classes with $t_i \in \{-1,1\}$ corresponding to an un-detected GRB and a detected GRB, respectively. A hyper-plane separating the two classes will satisfy $\bm{w} \cdot \bm{x} - b = 0$, where $\bm{w}$ and $b$ must be found by training on the data $\{\bm{x}\}$. If the classes are separable, we can place two parallel hyper-planes that separate the points and have no points between them in the ``margin''. This can be seen for a toy example in Figure~\ref{fig:SVMtoy1}. We describe these hyper-planes mathematically as
\begin{equation}
\bm{w} \cdot \bm{x} - b = \pm 1,
\end{equation}
Examples will lie on either side of the two planes such that
\begin{equation}
t_i \left( \bm{w} \cdot \bm{x}_i - b \right) \geq 1
\label{eq:SVMconstraint}
\end{equation}
for all samples, $\bm{x}_i$. As the samples are typically not separable, we introduce slack variables $\xi_i \geq 0$ that measure the misclassification of $\bm{x}_i$ by setting
\begin{equation}
t_i \left( \bm{w} \cdot \bm{x}_i - b \right) \geq 1 - \xi_i \,.
\label{eq:SVMconstraint_slack}
\end{equation}
We then seek to minimize
\begin{equation}
\label{eq:SVMcost}
\textrm{Cost}(\bm{w},\bm{\xi},b) = \frac{1}{2} \|\bm{w}\|^2 + C \sum_{i} \xi_i
\end{equation}
subject to the constraint in Equation~\ref{eq:SVMconstraint_slack}. The $C$ parameter is a penalty factor for misclassification and this optimization will face the trade-off between a smaller margin and smaller misclassification error. The cost function seeks to maximize the distance between the two hyper-planes at the margin edges, which is given by $2/\lvert\lvert \bm{w} \rvert\rvert$. This separation by hyper-plane is demonstrated for a toy example in Figure~\ref{fig:SVMtoy1}.

The two classes of points are generally not easily separated in the original parameter space of the problem. Therefore, we map the points into a higher-dimensional space where they may be more easily separated.
%A toy model of this is shown in Figure~\ref{fig:SVMtoy2}.
To make this a computationally tractable problem, we consider mappings such that the dot product between pairs of points may be easily computed in terms of the original variables by a kernel function, $k(\bm{x}_i,\bm{x}_j)$. Hyper-planes in the higher-dimensional space are defined as surfaces on which the kernel is constant. If the kernel is defined such that $k(\bm{x}_i,\bm{x}_j)$ decreases as the points $\bm{x}_i$ and $\bm{x}_j$ move away from one another, then the kernel is a measure of closeness. Thus, the sum of many kernels like this can be used to measure the proximity of a sample data point to data points in the two classes; this distance can then be used to classify the point into one class or the other. This mapping can result in a very convoluted hyper-plane separating the two sets of points -- this can accurately model the true classification boundary, but we must be careful not to overfit this to the training data.

In order to perform a non-linear separation, we employ a Gaussian kernel function (a.k.a. radial basis function),
\begin{equation}
\label{eq:gausskernel}
k(\bm{x}_i,\bm{x}_j) = \exp\left( -\gamma \| \bm{x}_i - \bm{x}_j \|^2 \right)
\end{equation}
where $\gamma$ is a tunable parameter reflecting the width of the Gaussian.

%\begin{figure}[htbp]
%\centering
%\includegraphics[width=0.9\columnwidth,keepaspectratio]{SVMtoy2}
%\caption{Toy demonstration of how a kernel can transform a non-linearly separable data set into a higher-dimensional linearly separable one. Image courtesy of Wikimedia Commons.}
%\label{fig:SVMtoy2}
%\end{figure}

A point is classified by which side of the learned hyper-plane it falls on, as determined (in our notation) by
\begin{equation}
\label{eq:SVMclass}
y(\bm{x}) = \mathrm{sign} \left( K(\bm{w},\bm{x})-b \right) \, ,
\end{equation}
where $K$ is the aggregate kernel function that is a linear combination of the individual kernel functions (closeness to each of the other points). Minimizing the cost given by Equation~\eqref{eq:SVMcost} under the constraint of Equation~\eqref{eq:SVMconstraint_slack} can be solved as a quadratic programming problem with the solution locally independent of all but a few data points, the ``support vectors'' of the model. These will be those samples closest to or on the margin in both classes and a weighted sum of distances from them will determine which class a new sample is in. Points that are not support vectors will have small or zero weight in the aggregate kernel function.

%In this study, we use the implementation of SVMs with a radial basis function kernel in the \texttt{kernlab} library~\cite{kernlabR} of the R statistical package. Cross-validation allows us to optimize for the hyper-parameters of the model, $\sigma$ and $C$. This function provides only discrete predictions, $y_{\mathrm{SVM}} \in \{-1,1\}$.
In this study, we use the implementation of SVMs in \texttt{scikit-learn}. A radial basis function is chosen and we perform $5$-fold cross-validation to optimize the hyper-parameters of the model, $\gamma$ and $C$. The model is also trained to allow for the prediction of continuous class probabilities, $y_{\mathrm{SVM}} \in [0,1]$\footnote{See the \texttt{scikit-learn} documentation for details on this procedure.}.

\subsection{Artificial Neural Networks}\label{sec:NNs}
Artificial neural networks are a machine learning method that is inspired by the function of a brain. A neural network (NN) consists of interconnected nodes, each of which processes information that it receives and passes this product on to other nodes via weighted connections. In a feed-forward NN, these nodes are organized into layers that pass information uniformly in a certain direction. The input layer passes information to an output layer via zero, one, or many ``hidden'' layers in between. Each node in the network performs a simple function, but their combined activity can model complex relationships. A useful introduction to NNs as well as their training and use can be found in~\cite{MacKayITILA}.% and a simple model is shown in Figure~\ref{fig:NNtoy}.

%\begin{figure}[htbp]
%\centering
%\includegraphics[width=0.65\columnwidth,keepaspectratio]{NNtoy}
%\caption{Toy model of a neural network. This NN has 3 input nodes, 4 hidden layer nodes, and 2 output nodes. Arrows indicate weighted connections between the nodes; biases are not shown. Image courtesy of Wikimedia Commons.}
%\label{fig:NNtoy}
%\end{figure}

A single node takes an input vector of activations $\bm{a} \in \Re^N$ and maps it to a scalar output $f(\bm{a}; \bm{w}, b)$ through
\begin{equation}
f(\bm{a}; \bm{w}, b) = g\left(b + \sum_{i=1}^N {w_i a_i}\right),
\label{eq:perceptron}
\end{equation}
where $\bm{w}$ and $b$ are the parameters of the node, called the ``weights'' and ``bias'', respectively. The function, $g$, is the activation function of the node; we use the sigmoid, linear, and rectified linear activation functions in this work.
\begin{equation}
\label{eq:activation}
g(z) = 
\begin{cases}
(1+\mathrm{e}^{-z})^{-1} & \textrm{sigmoid} \\
z & \textrm{linear} \\
\mathrm{max}\{0,z\} & \textrm{rectified linear}
\end{cases}
\end{equation}
The sigmoid and rectified linear activations are used for hidden layer nodes and the linear activation is used for the output layer nodes to obtain values in $(-\infty,\infty)$. This is then converted into a probability by the softmax transform given by
\begin{equation}
y_j(\bm{x};\bm{w},b) \rightarrow \frac{\exp\left( y_j(\bm{x};\bm{w},b) \right)}{\sum_{l=\{0,1\}} \exp\left( y_l(\bm{x};\bm{w},b) \right)}
\label{eq:softmax}
\end{equation}
where $j$ indexes over the output nodes. After the softmax, all output values are in $(0,1)$ and sum to $1$. We show here the case where there are only two output nodes for a binary classification problem; these values are degenerate, but the setup of one output node per class generalizes to the multi-class problem.

The weights and biases of all nodes in the network are the parameters that must be optimized with respect to the training data. The number of input nodes is the number of features given by the data. The two output nodes are the values in which are the probabilities that the input GRB features would result in detection or non-detection. In this work, we will take $y_{\mathrm{NN}}(\bm{x})$ to be the continuous probability given in the output node for the ``detection'' class. Thus, the output is the predicted probability that the given input GRB features correspond to a detected GRB.

The optimization algorithm seeks to minimize the cross-entropy of the predicted probabilities, given by
\begin{equation}
\textrm{Cost}(\bm{p}) = - \sum_{i} \sum_{k=\{0,1\}} t_{i,k} \log y_k(\bm{x}_i)
\label{eq:NNcost}
\end{equation}
where $\bm{p}$ is a parameter vector containing all of the weights and biases of the nodes in the NN. The index $i$ is over all data samples in the training set and the index $k$ is over the 2 output nodes corresponding to the non-detection and detection classes, respectively. $t_i = \{1,0\}$ for a non-detection and $t_i = \{0,1\}$ for a detection. This cost function pushes predicted probabilities toward their correct values with large penalties for incorrect predictions and is based in information theory. We take the value from the output node corresponding to the detection class as the probability that the input GRB, $\bm{x}$, is detected by \Sw{}, $y_{\rm NN}(\bm{x})$.

We use the \SN{}\footnote{\url{http://www.mrao.cam.ac.uk/software/skynet/}} algorithm~\citep{Graff2014} for training of the NN and refer the reader to that paper for more information on NNs, including the optimization function used, how the optimization is performed, and additional data processing that is performed. \SN{} provides an easy-to-use interface for training as well as an algorithm that will efficiently and consistently find the best fit NN parameters for the training data provided.

\subsection{Heuristics Used}\label{sec:heuristics}
For each model's optimal settings, we compute the accuracy of predictions using a na{\"i}ve probability threshold of $0.5$ for the output probability for the detection class; i.e. $y_m(\bm{x})$ for the different models, $m$. This is later found to be close to optimal. We also plot the receiver operating characteristic (ROC) curves for the classifiers as seen in Figure~\ref{fig:modelROCs} later. A ROC curve plots the true positive rate (a.k.a. recall) against the false positive rate. The F1-score is a useful metric for finding the optimal probability threshold to balance type I (false positive) and type II (false negative) errors. The F1-score takes values in $[0,1]$ and is maximized at the optimal probability threshold. These values are given by:
\begin{eqnarray}
\mathrm{TP} &=& \textrm{\# positives correctly labeled} \nonumber \\
\mathrm{TN} &=& \textrm{\# of negatives correctly labeled} \nonumber \\
\mathrm{FP} &=& \textrm{\# of negatives labeled as positive} \nonumber \\
\mathrm{FN} &=& \textrm{\# of positives labeled as negative} \nonumber \\
\mathrm{TPR} &=& \frac{\mathrm{TP}}{\mathrm{TP}+\mathrm{FN}} = \mathrm{recall} \nonumber \\
\mathrm{FPR} &=& \frac{\mathrm{FP}}{\mathrm{FP}+\mathrm{TN}} \nonumber \\
\mathrm{precision} &=& \frac{\mathrm{TP}}{\mathrm{TP}+\mathrm{FP}} \nonumber \\
\textrm{F1-score} &=& 2 \, \frac{\mathrm{precision} \times \mathrm{recall}}{\mathrm{precision} + \mathrm{recall}} \nonumber
\end{eqnarray}
where positives are detections and negatives are non-detections of GRBs. For a random classifier, the ROC will be a diagonal line from $(0,0)$ to $(1,1)$. Better classifiers will be above and to the left of this line. A common measure is the area under the curve (AUC), which is the integrated area under the ROC. Values closer to $1$ indicate better classifiers.

%We will also be comparing the redshift distributions of GRBs predicted as detected by each model to the true distribution in the validation data set. This comparison will be evaluated by the Kolmogorov-Smirnov (K-S) test, a non-parametric test of the equality of one-dimensional distributions. A larger p-value, $p\in[0,1]$, indicates a better fit between the true and predicted distributions (i.e. in favor of the null hypothesis that these two sets of samples are from the same underlying distribution). This is an important test, as we will be using the best model found to make predictions of GRB populations for Bayesian inference; an accurate reproduction of the true GRB redshift distribution is essential to this analysis.

In this study, we will use the ROC (with AUC) to find the MLA that best models the \Sw{} pipeline. We can then use the F1-score to identify the best probability threshold for declaring a detection from the predictions of the model.

\subsection{Cross-Validation}\label{sec:crossvalidation}
To measure the performance for each type of MLA, we perform hyper-parameter optimization over a range of settings for each. To properly compare these settings against each other, we perform \textit{cross-validation}. In this setup, with $5$ folds, we split the data into $5$ random subsets of equal size. In training, we train $5$ models for each setting, using $4$ of the $5$ during fitting and then evaluating the model on the left-out set. Thus we make predictions on the entire set but without having trained the model on the data it was predicting (this would lead to over-fitting).

Once the optimal model settings are found for each MLA, the entire data set is used to re-train a model with those values. This model is then evaluated on the left-out validation data set so as to compare it with the other MLAs. This latter test is much more stringent, as the evaluation data is from different populations than what was used in training. This better reflects how the ML model will be used in practice and is used to pick a MLA and model fit for use in Bayesian parameter estimation (Section~\ref{sec:bayesinf}).

%++++++++++++++++++++++++++++++++++++++++++++++++++++++++++++++++++++++
%
%  MACHINE LEARNING RESULTS
%
%++++++++++++++++++++++++++++++++++++++++++++++++++++++++++++++++++++++

\section{Machine Learning Results}\label{sec:results}
In this section we present the details of the MLA model fitting we performed. We describe the data set used for training and validation followed by results from hyper-parameter optimization searches performed for each classifier. The hyper-parameter optimization uses only the training data and evaluates different settings with cross-validation as described in Section~\ref{sec:crossvalidation}. Once we obtain optimal settings for each MLA, we evaluate the models on a validation data set (separate from the training data) for final performance measurement and comparison.

\subsection{Training Data Used}\label{sec:train data}
\begin{figure}
\centering
\includegraphics[width=0.8\columnwidth]{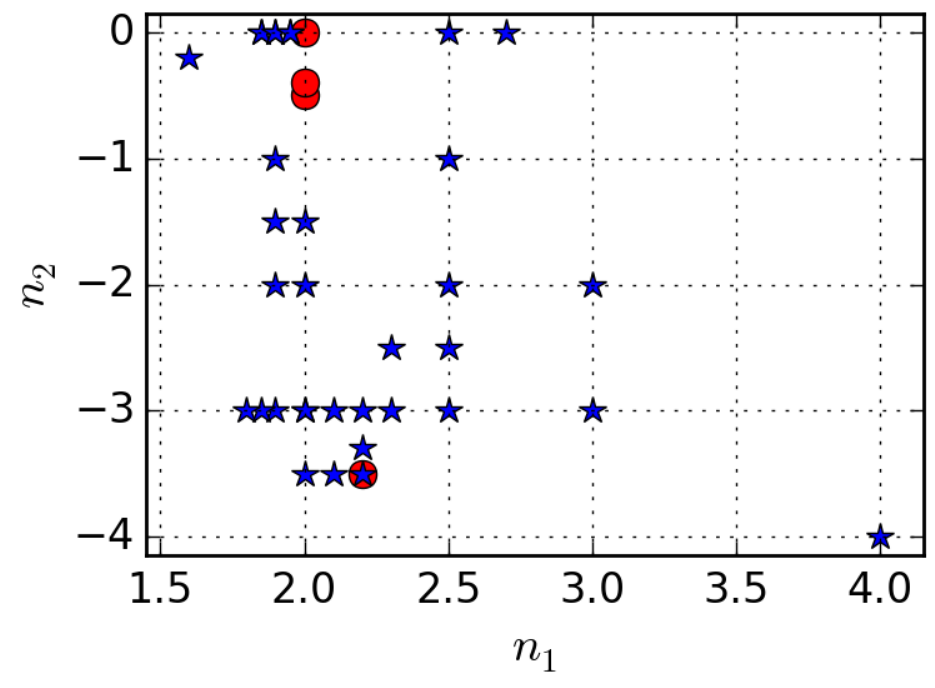}
\caption{Values of the parameters for the redshift distribution function for sample GRB populations used to train ML models. Blue stars were used in training and optimization, red circles were used for final evaluation.}
\label{fig:train_samp_distr}
\end{figure}

The data used in this analysis was generated by simulations of the \Sw{} pipeline -- as described in Section~\ref{sec:swift_pipe} -- for different settings of the GRB redshift and luminosity distribution functions (Equations 2 and 3 in~\cite{Lien2014} and reproduced below).
\begin{align}
R_{\mathrm{GRB}}(z) &= n_0
\begin{cases}
(1+z)^{n_1} & z \leq z_1 \\
(1+z_1)^{n_1-n_2}(1+z)^{n_2} & z > z_1
\end{cases} 
\label{eq:redshiftfctn} \\
\phi(L) &= \frac{dN}{dL} =
\begin{cases}
\left(\frac{L}{L_{\star}}\right)^x & L \leq L_{\star} \\
\left(\frac{L}{L_{\star}}\right)^y & L > L_{\star}
\end{cases}
\label{eq:luminosityfctn}
\end{align}
$R_{\mathrm{GRB}}(z)$ is the comoving GRB rate, with units of $\mathrm{Gpc}^{-3}\mathrm{yr}^{-1}$. In these data sets, the luminosity distribution function was held constant with $x=-0.65$, $y=-3.00$, and $L_{\star}=10^{52.05} \, \mathrm{erg}/\mathrm{s}$. Additionally, the break in the redshift distribution was also held constant at $z_1=3.60$. Therefore, we only varied values of $n_1$ and $n_2$ ($n_0$ is ignored for the purpose of generating training data as it is only a normalization parameter). In total, 38 datasets are combined for use in training. These datasets were originally generated for~\cite{Lien2014} and do not cover the space systematically. We use 34 of the 38 data sets for training models, including optimization of hyper-parameters; each of these contains $\sim4000$ samples. The final 4, which contain $\sim10000$ samples each, are set aside for evaluating the final model from each MLA as the validation data. The distribution of parameters for each of these data sets is shown in Figure~\ref{fig:train_samp_distr}.

We used this data for training as it was generated around the best-fit values from~\cite{Lien2014} for the real \Sw{} GRB redshift measurements of~\cite{Fynbo2009}. In the end, our goal is to fit the GRB rate model to these same observations.

A total of $15$ parameters are taken from each simulated GRB in order to determine whether or not the GRB was detected by \Sw{}. These are summarized in Table~\ref{tab:GRBparams}. These are used for classification of GRBs by MLAs. The target value is given by the trigger\_index, which is $0$ for GRBs that are not detected by the \Sw{} algorithm and $1$ for those that are detected.

\setlength{\tabcolsep}{8pt}
\begin{table}
\begin{center}
\begin{tabular}{c | l}
\textbf{Parameter} & \textbf{Description} \\
\hline
\hline
$\log_{10}(L)$ & luminosity of the GRB \\
z & redshift \\
$r$ & distance from center of detector grid of peak \\
$\phi$ & azimuthal angle in detector grid of peak \\
bin\_size\_emit & source time bin size \\
$\alpha$ & Band function parameter \\
$\beta$ & Band function parameter \\
$\log_{10}(E_{\mathrm{peak}})$ & peak of the energy spectrum of the GRB \\
bgd\_15-25keV & background count rate in $15$--$25$keV band \\
bgd\_15-50keV & background count rate in $15$--$50$keV band \\
bgd\_25-100keV & background count rate in $25$--$100$keV band \\
bgd\_50-350keV & background count rate in $50$--$350$keV band \\
$\theta$ & incoming angle of GRB \\
$\log_{10}(\Phi)$ & incident flux of GRB \\
ndet & number of active detector pixels (constant) \\
\hline
trigger\_index & 0 for non-detections and 1 for detections
\end{tabular}
\end{center}
\caption{Parameters describing each simulated GRB. There are $15$ inputs and the output class label. See Equation 4 in~\cite{Lien2014} for details of $\alpha$, $\beta$, and $\log(E_{\mathrm{peak}})$.}
\label{tab:GRBparams}
\end{table}
\setlength{\tabcolsep}{6pt}

\begin{figure}
\centering
\includegraphics[width=0.95\columnwidth]{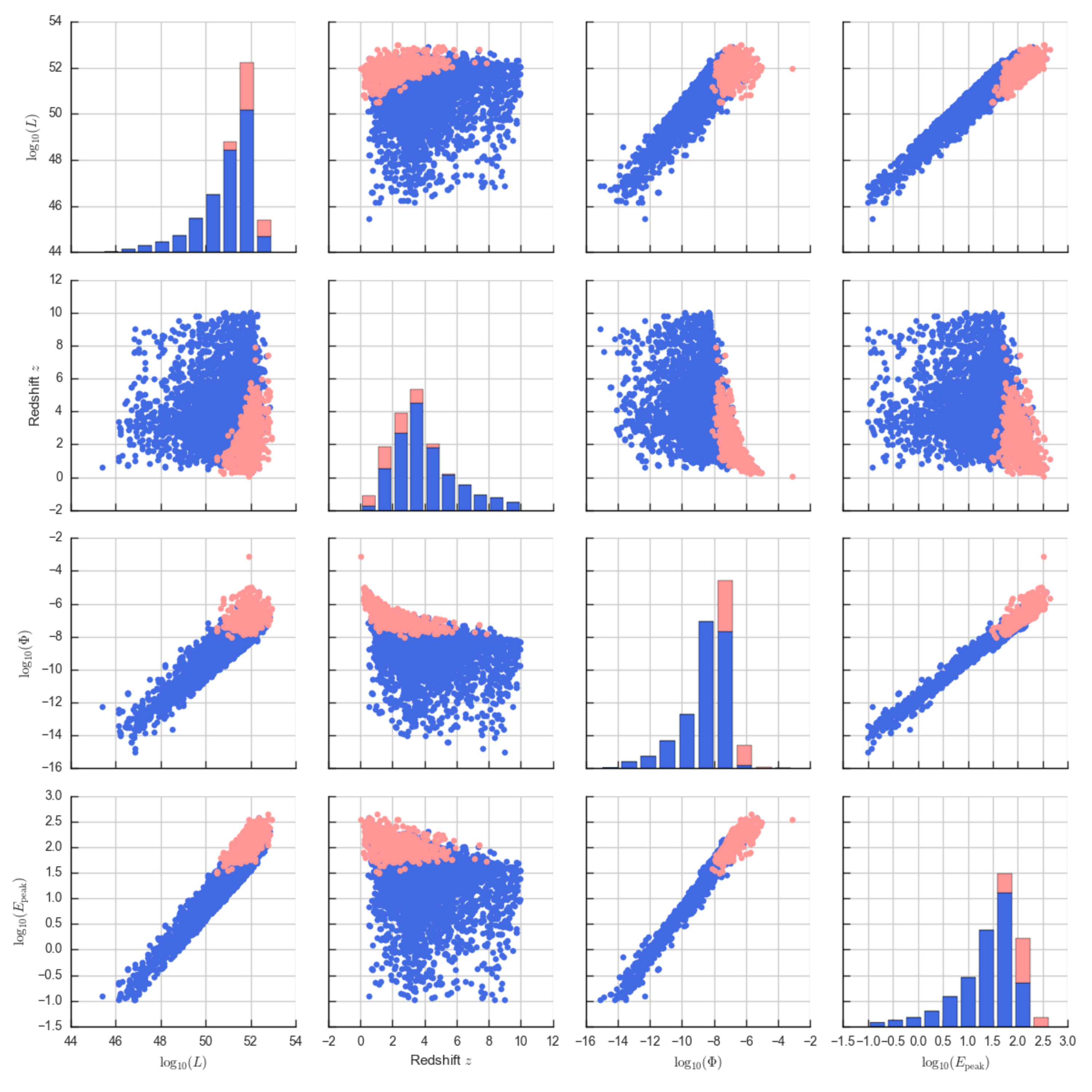}
\caption{Pair-wise scatterplot of a few of the most significant parameters in determining detection. A random subset of $5000$ points from the training data set is shown. GRBs that are detected are indicated by light red points, non-detected ones are blue.}
\label{fig:pairplot}
\end{figure}

A pair-wise plot of a few of the most significant parameters in determining detection is shown in Figure~\ref{fig:pairplot}. Lighter points are GRBs that are detected by \Sw{} in the trigger simulator~\citep{Lien2014} while darker ones are undetected GRBs. This plot shows a random subset of $5000$ points from the entire training data set.

To determine how much training data is required, we evaluated the learning curve for the random forest classifier. This plots the prediction accuracies, computed using $5$-fold cross-validation, as a function of the size of the training data.
The ``training data'' in this case is the \rfrac{4}{5} of the data used for fitting the model and the ``test data'' is the \rfrac{1}{5} left out for evaluation.
The learning curve was done after finding the optimal RF settings in Section~\ref{sec:RFresults} as a check. We thus examined if use of the entire data set benefits model fitting significantly. The data was randomly shuffled before performing this test.

\begin{figure}[htbp]
\centering
%\captionsetup[subfigure]{labelformat=empty}
%\begin{subfigure}{\columnwidth}
\includegraphics[width=0.9\columnwidth]{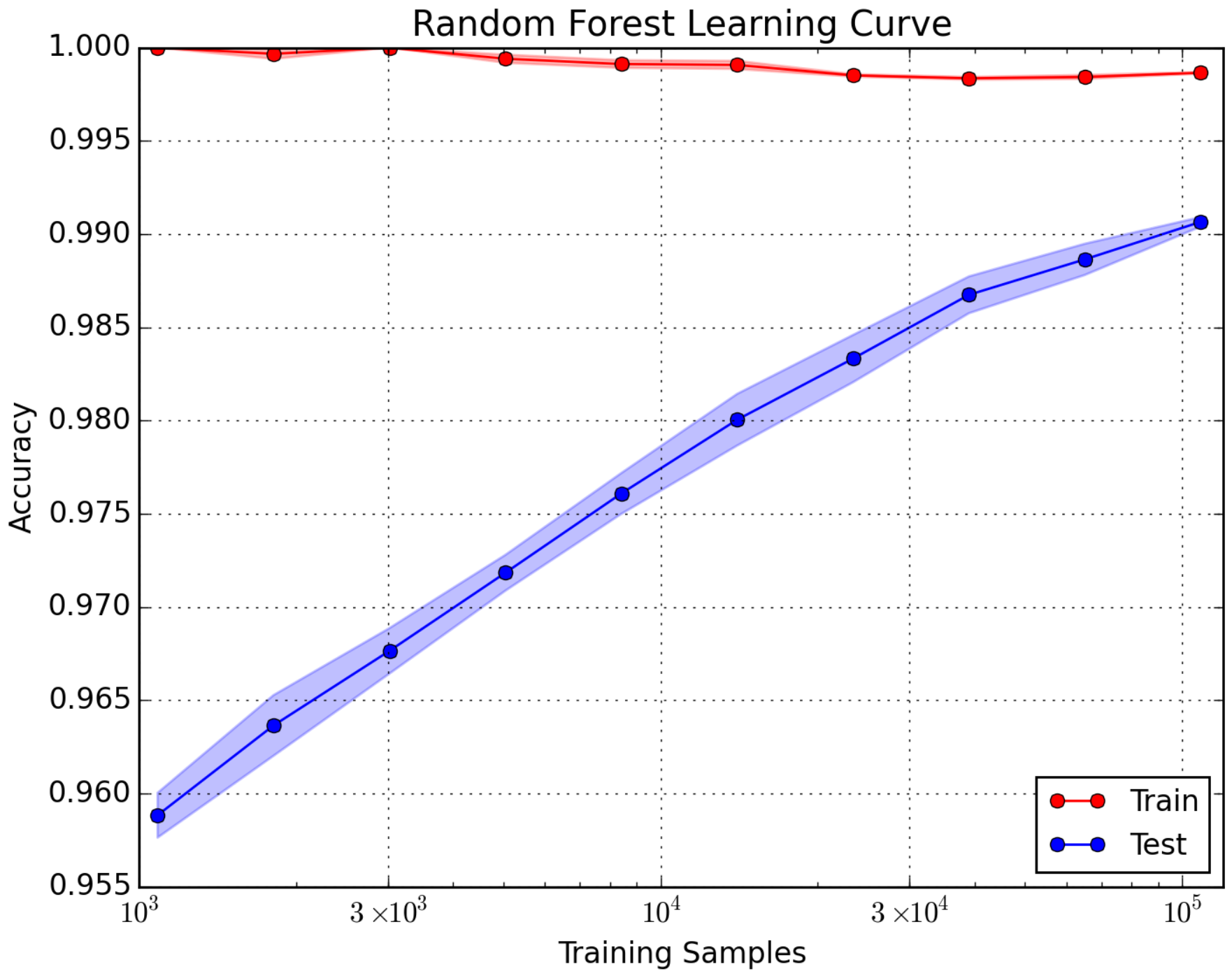}
%\end{subfigure}
%\begin{subfigure}{\columnwidth}
%\includegraphics[width=\columnwidth]{ABlearncurve}
%\end{subfigure}
\caption{Learning curve for the random forest classifier. The training data set accuracy is fairly constant above $3\times10^4$ samples but the test accuracy continues to increase with the number of data points.}
\label{fig:learning_curves}
\end{figure}

The resulting learning curve is shown in Figure~\ref{fig:learning_curves}. For small sample sizes, there is overfitting of the training data that begins to flatten out by $3\times10^4$ samples. The accuracy of the test set continues to increase as we add more data points, meaning that more data improves the generalizability of the model. Therefore, in all subsequent training we will use the entire data set for fitting a model; using fewer points would increase the bias of subsequent predictions.

\subsection{Random Forest}\label{sec:RFresults}
\begin{figure}[htbp]
\centering
\includegraphics[width=\columnwidth]{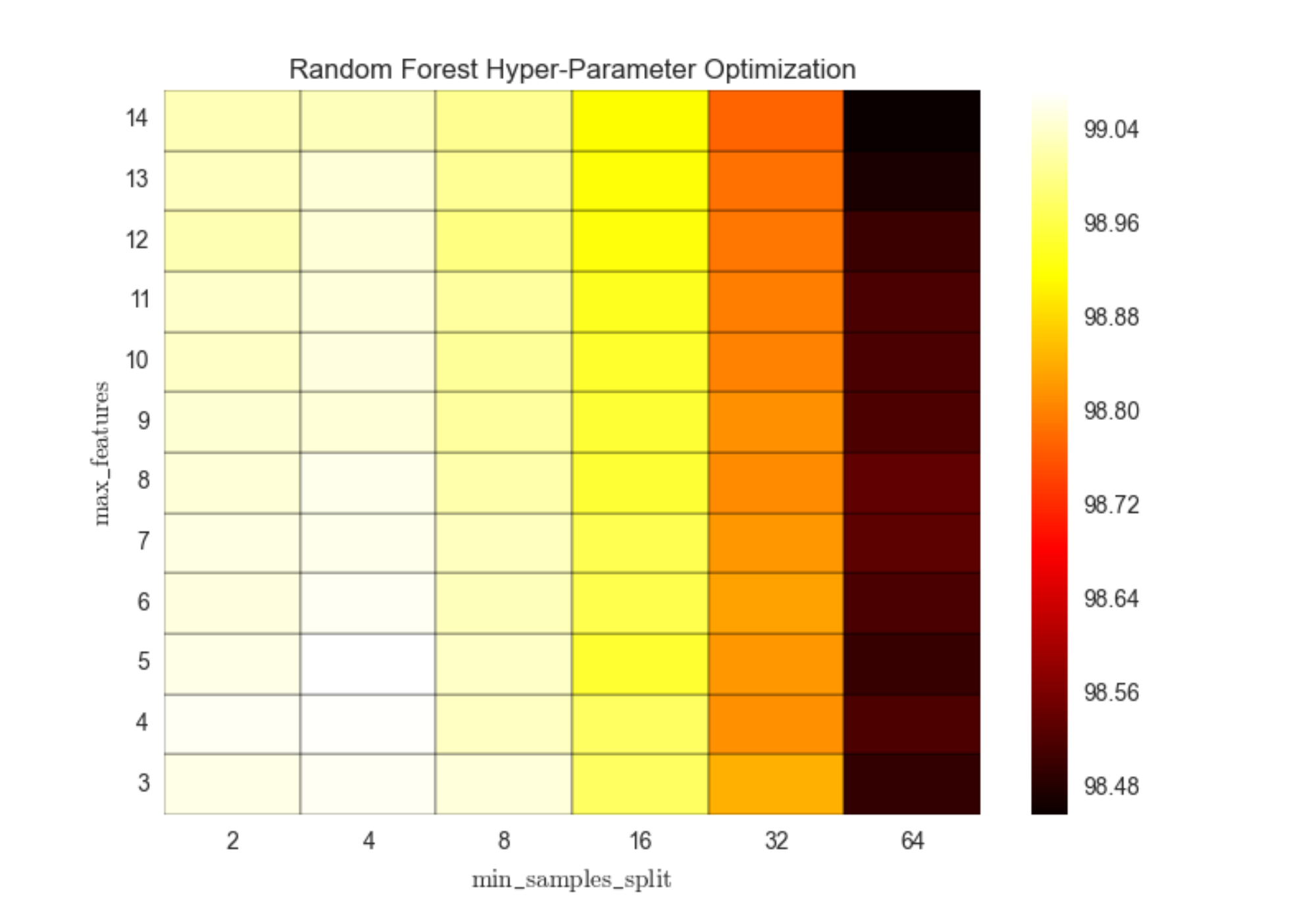}
\caption{Test set accuracy for random forest classifier hyper-parameters. The optimal value is $(4,5)$. It is clear that \mss{} is a much stronger influence on over-fitting to the training data.}
\label{fig:RFgridsearch}
\end{figure}

The random forest model was optimized for combinations of the \mss{} and \mf{} parameters. These govern the minimum number of samples needed to perform a branching split and the number of features considered at each split, respectively. Choices for each are as follows\footnote{The values for \mss{} go from the absolute minimum, $2$, to a significantly larger value where we see degraded performance by powers of $2$. The choices for \mf{} vary from a low number (minimum is $1$) to the maximum value that doesn't consider every parameter at each split and thus would have no randomness.}:
\begin{align}
\mss{} &\in \{2, 4, 8, 16, 32, 64\} \nonumber \\
\mf{} &\in \{3,4,5,6,7,8,9,10,11,12,13,14\} \nonumber
\end{align}
Forests were trained with $500$ trees, using the Gini impurity for deciding the optimal split at each branching point and with no limit on the number of branches before reaching a leaf. The $5$-fold cross-validation evaluates the test accuracy for each pairwise combination of the parameters; the set with the highest test accuracy is the optimal model. The optimal parameters found were $\mss{}=4$ and $\mf{}=5$, however, it can be seen in Figure~\ref{fig:RFgridsearch} that there is very little variation in accuracy with regard to the value of \mf{}. The minimum number of samples required to make a split is the dominant factor for improving the accuracy, where smaller values that naturally fine-tune the model further obtain better accuracy on the test set as well. The overall range in test set accuracy is not large and the worst model hyper-parameters still achieve accuracy $>98\%$. The preference for lower values in \mf{} can be understood as increasing variability between trees in the forest and thus minimizing over-fitting.

Using the optimal model, we perform predictions on the validation data set. With a na{\"i}ve threshold of $0.5$ on the output probability for the detection class for declaring a GRB detected, these predictions have an accuracy of $97.5\%$. This is lower than the test accuracy obtained earlier as the test samples were from the same distribution as the training ones while this validation data presents new distributions. The ROC for this classifier is shown with the others in Figure~\ref{fig:modelROCs} and has an $\mathrm{AUC} = 0.9935$.
%The K-S test p-value, averaged between the four distributions in the validation data set, is $1.000$ (when rounded to $3$ decimal places).
Analysis of the F1-score found no significant difference between the optimal probability threshold and the na\"{i}ve threshold of $0.5$.

\subsection{AdaBoost}\label{sec:ABresults}
The AdaBoost model was optimized for combinations of the \nes{} and \lr{} parameters. The former describes the number of `weak learners' (decision trees) fit in each ensemble model and the latter describes the rate for adjusting the weighting of the weak learners as each is added to the ensemble. Settings for the individual decision trees were chosen to match those found as optimal for the random forest classifier, with $\mss{}=4$ and $\mf{}=5$. Choices for each were as follows\footnote{The \nes{} range was determined by having enough trees for refined probability estimates while not needing more than the RF model. The range for the \lr{} parameter goes from a large value, $1$, down to a small rate; we did not test smaller values as all models achieved very similar performance with each other and with the best RF model.}
\begin{align}
\nes{} &\in \{100, 200, 300, 400, 500\} \nonumber \\
\log_{10}(\lr{}) &\in \{-3, -2.5, -2, -1.5, -1, -0.5, 0\} \nonumber
\end{align}
The $5$-fold cross-validation found that the optimal parameters are $(100,0.001)$. However, the range in test set accuracies is extremely small, varying only between $99.01\%$ and $99.05\%$. Therefore, any of these models would be nearly equally accurate.
%We do, however, see a small trend towards higher test accuracy with lower learning rates, which down-weights later classifiers more; this is consistent with those later classifiers becoming more specialized to the training data and thus potentially over-fitting and introducing bias.
\begin{figure}[htbp]
\centering
\includegraphics[width=\columnwidth]{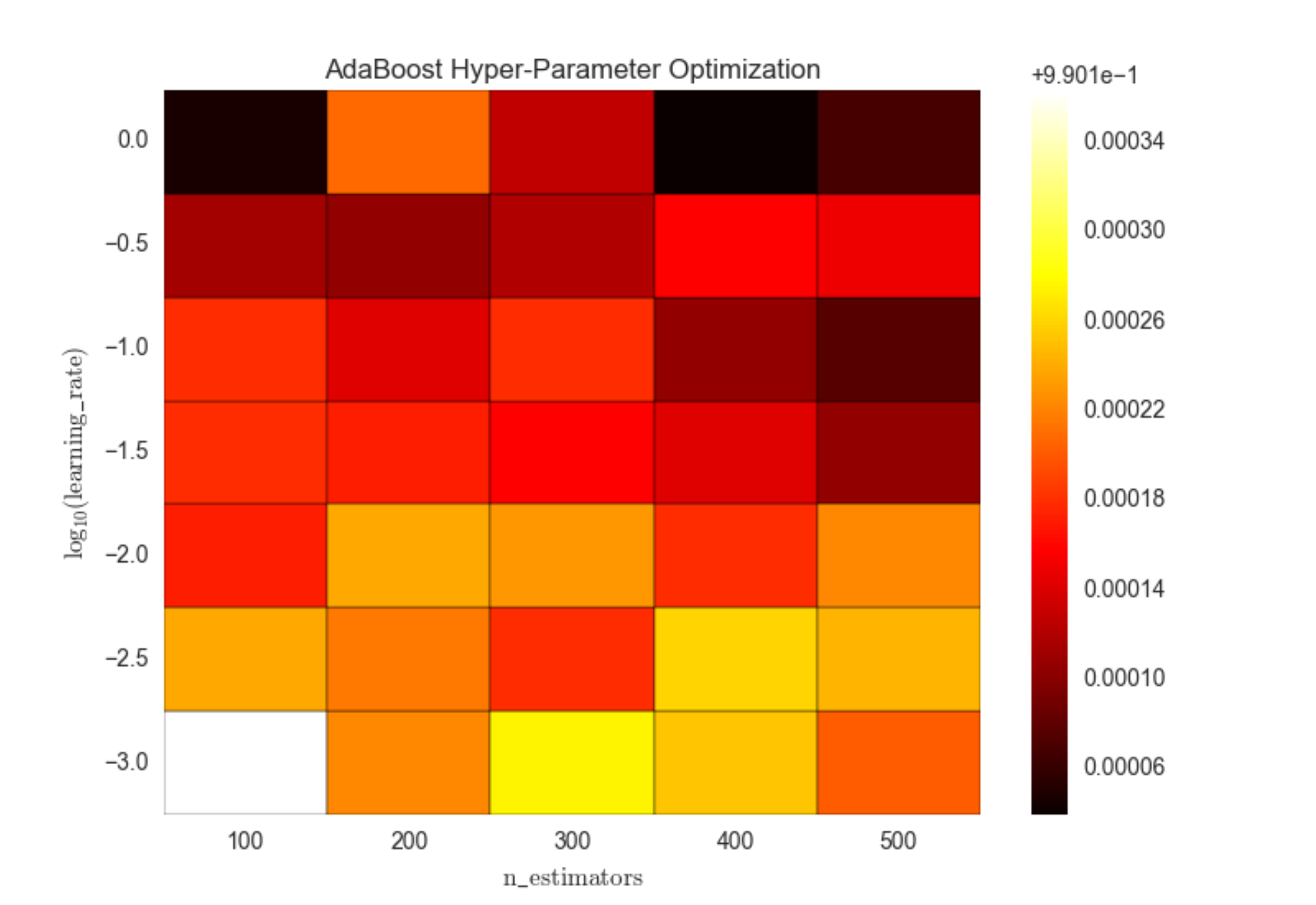}
\caption{Test set accuracy for AdaBoost classifier hyper-parameters. The optimal value is $(100,0.001)$. All options are very close to each other, ranging only from $99.01\%$ to $99.05\%$ in accuracy.}
\label{fig:ABgridsearch}
\end{figure}

Using the optimal model, we perform predictions on the validation data set. With a na{\"i}ve threshold of $0.5$ on the output probability for the detection class for declaring a GRB detected, these predictions have an accuracy of $97.4\%$. The ROC for this classifier is shown with the others in Figure~\ref{fig:modelROCs} and has an $\mathrm{AUC} = 0.9921$.
%The K-S test p-value, averaged between the four distributions in the validation data set, is $1.000$ (when rounded to $3$ decimal places).
Analysis of the F1-score found no significant difference between the optimal probability threshold and the na\"{i}ve threshold of $0.5$.

\subsection{Support Vector Machines}\label{sec:SVMresults}
The support vector machine model was trained using a Gaussian (radial basis function) kernel, as described in Equation~\eqref{eq:gausskernel}. The input data values were all scaled to have zero mean and unit variance, so as to prevent undue bias in the kernel's distance measure. As errors in the predictions are allowed, there are thus two hyper-parameters to optimize, the penalty factor for errors, $C$, and the tunable parameter for the width of the Gaussian, $\gamma$. Choices examined for these were (after first searching over a larger grid with coarser spacing):
\begin{align}
%\log_{10}(C) &\in \{-2,-1,0,1,2,3,4,5,6\} \nonumber \\
%\log_{10}(\gamma) &\in \{-5,-4,-3,-2,-1,0,1,2,3\} \nonumber
\log_{10}(C) &\in \{1.25,1.5,1.75,2,2.25,2.5,2.75,3\} \nonumber \\
\log_{10}(\gamma) &\in \{-1,-0.75,-0.5,-0.25,0,0.25,0.5,0.75,1\} \nonumber
\end{align}
$5$-fold cross validation found optimal parameters of $(C,\gamma)=(10^{2.25},1)$ with a test set accuracy of $99.0\%$. For smaller values of $C$, there is a much more limited range in $\gamma$ that gives comparable results, if at all.
\begin{figure}[htbp]
\centering
\includegraphics[width=\columnwidth]{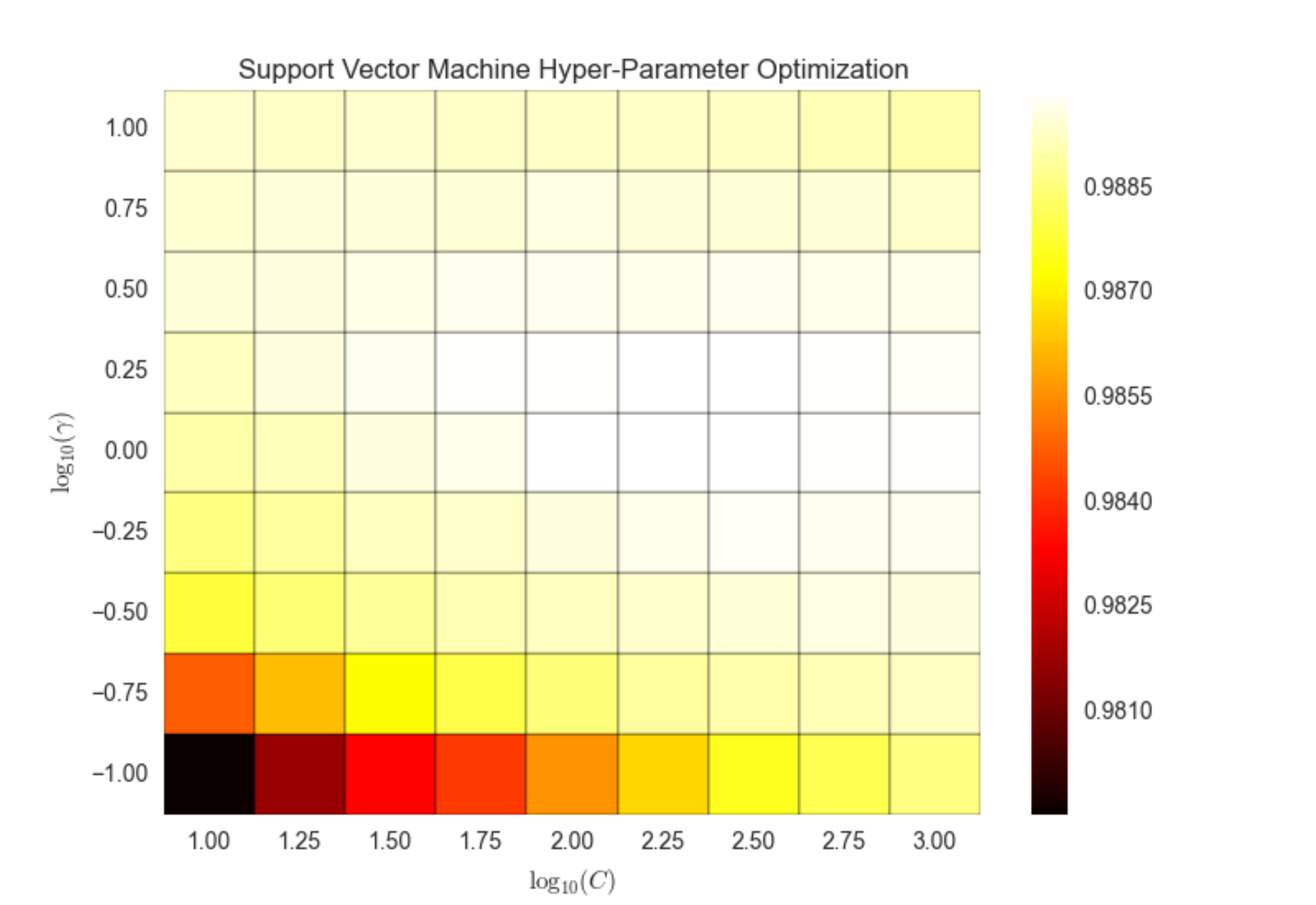}
\caption{Test set accuracy for support vector machine classifier hyper-parameters. The optimal value is $(C,\gamma)=(10^{2.25},1)$.}%The color bar has been adjusted so as to better show differences at higher accuracies (values $\leq0.9$ are all black).
\label{fig:SVMgridsearch}
\end{figure}

Using the optimal model and a $0.5$ probability threshold for classification as a detection, the SVM has a prediction accuracy of $94.5\%$. The ROC for this classifier is shown in Figure~\ref{fig:modelROCs} and has an $\mathrm{AUC}=0.9348$.
%The K-S test p-value, averaged between the four distributions in the validation data set, is $0.223$.
From all of these measures it is clear that the SVM model, in this scenario, does not generalize as well as the decision tree ensemble methods (RF and AdaBoost).

\subsection{Neural Networks}\label{sec:NNresults}
Using \SN{}, we trained several neural network architectures using either the sigmoid or rectified linear activation function for the hidden layer nodes. Training NNs is much more computationally expensive than any of the other models, despite the efficiencies in the training algorithm. Therefore, the size of our NN models (in both number and width of hidden layers) as well as the time spent training them, is limited. For each architecture\footnote{The architecture is given by X or X+Y, the former indicating a single hidden layer with X nodes and the latter indicating two hidden layers with X and Y nodes, respectively.} we employed $5$-fold cross validation in order to asses its performance. We report in Table~\ref{tab:NNtrainresults} the test set accuracies for each of the networks trained. They are all similar and are getting close to the $99\%$ achieved by the previous MLAs. It is possible that more complex networks would achieve this level of accuracy.
\setlength{\tabcolsep}{8pt}
\begin{table}
\centering
\begin{tabular}{cc|c}
\textbf{Hidden Layers} & \textbf{Activation} & \textbf{Test Accuracy} \\
\hline
\hline
\multirow{2}{*}{25} & sigmoid & 97.89 \\
 & rectified & 97.25 \\
\hline
\multirow{2}{*}{50} & sigmoid & 98.33 \\
 & rectified & 97.57 \\
\hline
\multirow{2}{*}{100} & sigmoid & 98.47 \\
 & rectified & 98.00 \\
\hline
\multirow{2}{*}{1000} & sigmoid & 97.49 \\
 & rectified & 98.28 \\
\hline
\multirow{2}{*}{25+25} & sigmoid & 98.33 \\
 & rectified & 97.65 \\
\hline
\multirow{2}{*}{50+50} & sigmoid & 98.73 \\
 & rectified & 98.16 \\
\hline
\multirow{2}{*}{100+30} & sigmoid & 98.47 \\
 & rectified & 98.27 \\
\hline
\multirow{2}{*}{100+50} & sigmoid & 98.64 \\
 & rectified & 98.41 \\
\hline
\multirow{2}{*}{100+100} & sigmoid & 97.95 \\
 & rectified & 98.35
\end{tabular}
\caption{Test set accuracy from $5$-fold cross-validation from the training of neural networks with \SN{}. The activation functions are given in Equation~\eqref{eq:activation}.}
\label{tab:NNtrainresults}
\end{table}
\setlength{\tabcolsep}{6pt}

\begin{figure}[htbp]
\centering
\includegraphics[width=0.9\columnwidth]{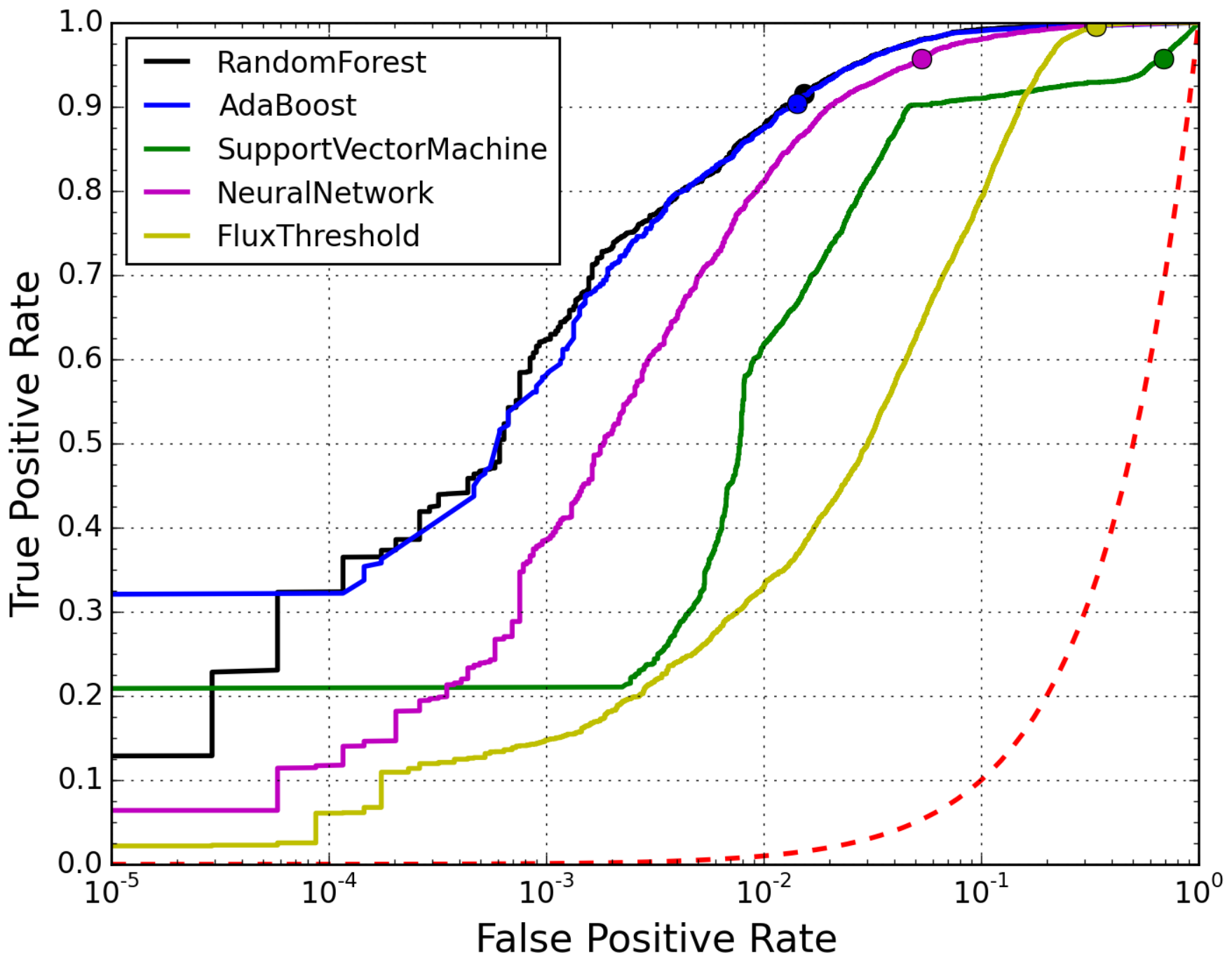}
\caption{Receiver operating characteristic (ROC) curves for the classifiers. A dot is placed at the values for the optimal probability threshold found for each classifier. The ROC curve of a random classifier is shown in a dashed red line. A logarithmic scale for the x-axis is used to display the differences in the ROC curves.}
\label{fig:modelROCs}
\end{figure}

Due to the constraints on training, we consider the NN architecture with highest average test accuracy, considering both activation functions: the 100+50 architecture of hidden layers. This is retrained on the entire data set with both activation functions and we find that the optimal model has hidden layers of 100+50 with the rectified linear unit activation function. We use this NN to make predictions on the validation data set with a na\"{i}ve probability threshold of $0.5$. This yields and accuracy of $96.9\%$. The ROC curve for this NN is shown in Figure~\ref{fig:modelROCs} and has an $\mathrm{AUC}=0.989$.
%The K-S test p-value, averaged between the four distributions in the validation data set, is $0.999$.
Analysis of the F1-score found no significant difference between the optimal probability threshold and the na\"{i}ve threshold of $0.5$.

\subsection{Summary of Results}\label{sec:results_summary}
Here we summarize the results for the optimal model returned by each MLA. The accuracy, AUC, and optimal F1-score are all reported in Table~\ref{tab:model_results}. We also include in this comparison the use of a constant cut in GRB flux; GRBs with flux greater than a threshold value will be labeled as detected and those with lower flux are non-detections. Varying this flux threshold produces a ROC and we find an optimal cut at $\log_{10}(\Phi) = -7.243 \,\,\mathrm{erg/s/cm^2}$ (based on the F1-score) for which we measure the accuracy. It is clear that all ML classifiers except SVMs significantly outperform a flux threshold; SVMs still outperform a flux cut at optimal settings.

\begin{table}
\centering
\begin{tabular}{c | c | ccc}
\textbf{Classifier} & \textbf{Threshold} & \textbf{Accuracy} & \textbf{AUC} & \textbf{F1-score} \\
\hline
\hline
Random Forest & 0.449 & 0.975 & 0.994 & 0.912\\
AdaBoost & 0.362 & 0.975 & 0.992 & 0.910\\
Neural Net & 0.459 & 0.969 & 0.989 & 0.890\\
SVM & 0.028 & 0.947 & 0.935 & 0.824\\
Flux & -7.243 & 0.896 & 0.945 & 0.663
\end{tabular}
\caption{Results for measuring the performance of the classifiers trained in this study. The accuracy on the validation data, the area under the ROC curve, and the optimal F1-score are reported. The threshold values are probabilities for all models except the flux cut, which uses the $\log_{10}(\Phi)$.}
\label{tab:model_results}
\end{table}

From this analysis, we see that the RF and AdaBoost classifiers performed the best in classification task. NNs were very close behind, with SVMs performing the worst among the MLAs\footnote{It should be noted that this is not a comment on the general performance of the MLAs, merely how well they performed on this task with this data set.}.

%++++++++++++++++++++++++++++++++++++++++++++++++++++++++++++++++++++++
%
%  BAYESIAN PE
%
%++++++++++++++++++++++++++++++++++++++++++++++++++++++++++++++++++++++

\section{Use of Accelerated Pipeline for Bayesian Inference}\label{sec:bayesinf}
%In order to simplify the integration of the learned model and the code bases that produce the GRB population and perform Bayesian sampling, we will use the NN model from \SN{}. We expect the loss in precision -- relative to using RFs or AdaBoost -- to be minimal.

%In the following subsection, we first describe the GRB source population model we will be fitting, including its free/fixed parameters and the chosen Bayesian priors. We then perform the Bayesian inference and present results from the posterior probability distribution of the model's free parameters. Lastly, we discuss the greatly reduced computational cost of performing this kind of analysis.

Here we demonstrate the use of the trained ML models in accelerating Bayesian inference, namely fitting the intrinsic redshift distribution of GRBs. We do so with best-fit random forest, AdaBoost, and \SN{} NN models, these being the most accurate.

\subsection{Likelihood Function}\label{sec:bayesLogL}
We first consider how we will evaluate the fit of a model to a set of GRB redshift observations, a.k.a. ``the data''. If we bin the observations to obtain a redshift density, then in each bin (with central redshift $z_i$) there will be an observed number of GRBs, $N_{\rm obs}(z_i)$. There is also an expected number of intrinsic GRBs occurring in \Sw{}'s field of view during the observation time in each redshift bin given by
\begin{equation}
N_{\rm int}(z_i) = \frac{4 \pi}{6} \Delta t_{\rm obs} R_{{\rm GRB};dz}(z_i) dz,
\label{eq:intrinsic_bin_GRBs}
\end{equation}
where
\begin{equation}
R_{{\rm GRB}; dz}(z) = \frac{R_{\rm GRB}(z)}{1+z} \frac{dV_{\rm comov}}{d\Omega dz}.
\label{eq:redshift_obs}
\end{equation}
$R_{{\rm GRB}; dz}(z)$ is the observed GRB rate that accounts for time dilation and the comoving volume in addition to the comoving rate, $R_{{\rm GRB}}(z)$. The $\rfrac{4\pi}{6}$ factor introduced here reflects that \Sw{} observes only a sixth of the entire sky and $\Delta t_{\rm obs}$ reflects the fraction of time (per year) that \Sw{} is observing; this is taken as $\Delta t_{\rm obs} \approx 0.8$ as calculated from related \Sw{} log data. $V_{\rm comov}$ is the cosmological co-moving volume and $\Omega$ is the subtended sky angle.

Not all GRBs occurring in \Sw{}'s field of view will be detected, however; this is taken into account by the extra factor, $F_{\rm det}(z)$. This is the fraction of GRBs at redshift $z$ that are detected by \Sw{} and is further discussed in Section~\ref{sec:detection_fraction}. Including this factor gives us the expected number of observed GRBs in each bin,
\begin{equation}
N_{\rm exp}(z_i) = \frac{4 \pi}{6} \Delta t_{\rm obs} R_{{\rm GRB};dz}(z_i) F_{\rm det}(z_i) dz.
\label{eq:expected_bin_GRBs}
\end{equation}

The probability, then, of observing $N_{\rm obs}(z_i)$ GRBs when $N_{\rm exp}(z_i)$ are expected is given by the Poisson distribution. The bins can be treated as independent, so for $K$ bins we can multiply their probabilities.
\begin{align}
\Pr(\{N_{\rm obs}(z_i)\};\{N_{\rm exp}(z_i)\}) &= \prod_{i=1}^{K} \Pr(N_{\rm obs}(z_i);N_{\rm exp}(z_i)) \notag \\
&= \prod_{i=1}^{K} \frac{N_{\rm exp}(z_i)^{N_{\rm obs}(z_i)} e^{-N_{\rm exp}(z_i)}}{N_{\rm obs}(z_i)!}
\end{align}
The log-likelihood is therefore the log of this probability,
\begin{align}
\mathcal{L}(\vec{n}) &= \log\left(\Pr(N_{\rm obs}(z_i);N_{\rm exp}(z_i))\right) \notag \\
&= \sum_{i=1}^{K} N_{\rm obs}(z_i) \log(N_{\rm exp}(z_i)) - N_{\rm exp}(z_i) - \log(N_{\rm obs}(z_i)!)
\label{eq:binned_loglike}
\end{align}
where $\vec{n} = \{n_0, n_1, n_2, z_1, x, y, L_{\ast}\}$ is the set of model parameters that let us obtain $N_{\rm exp}(z_i)$, which is really $N_{\rm exp}(z_i \vert \vec{n})$.

In the limit of a large number of bins, each bin will contain either $0$ or $1$ detected GRBs so $N_{\rm obs}(z_i)! = 1 \Rightarrow \log(N_{\rm obs}(z_i)!) = 0$. We can also split terms and rewrite Eq~\eqref{eq:binned_loglike} as
\begin{align}
\mathcal{L}(\vec{n}) &= \sum_{i=1}^{K} \left[N_{\rm obs}(z_i) \log(N_{\rm exp}(z_i))\right] - \sum_{i=1}^{K} N_{\rm exp}(z_i) \notag \\
&= -N_{\rm exp} + \sum_{\{i\}_{\rm det}} \log(N_{\rm exp}(z_i))
\end{align}
where $\{i\}_{\rm det}$ are those bins with a detection. We can perform this calculation in the limit of infinite bins, essentially a continuous measurement. $N_{\rm exp}$ is the integrated expected rate of observations given by
\begin{equation}
N_{\rm exp} = \int_0^{10} N_{\rm exp}(z) dz.
\end{equation}

This likelihood is the same as the $C$-statistic derived in~\cite{Cash1979} in the un-binned limit (see Equation 7 therein, where $C = -2 \mathcal{L}$). This likelihood function is also equivalent to that of~\cite{Stevenson2015}, which compares discrete intrinsic population models for binary black hole mergers as observed by advanced LIGO and Virgo using the observed mass distribution, if the latter is taken to the same limit of infinite bins of infinitesimal width. This is particularly notable as~\cite{Stevenson2015} uses a Poisson probability for the total number of detections multiplied by a multinomial distribution describing the fractional distribution of detections among bins in mass space. 

\subsubsection{Detection Fraction}\label{sec:detection_fraction}
\begin{figure}[htbp]
\centering
\includegraphics[width=0.9\columnwidth]{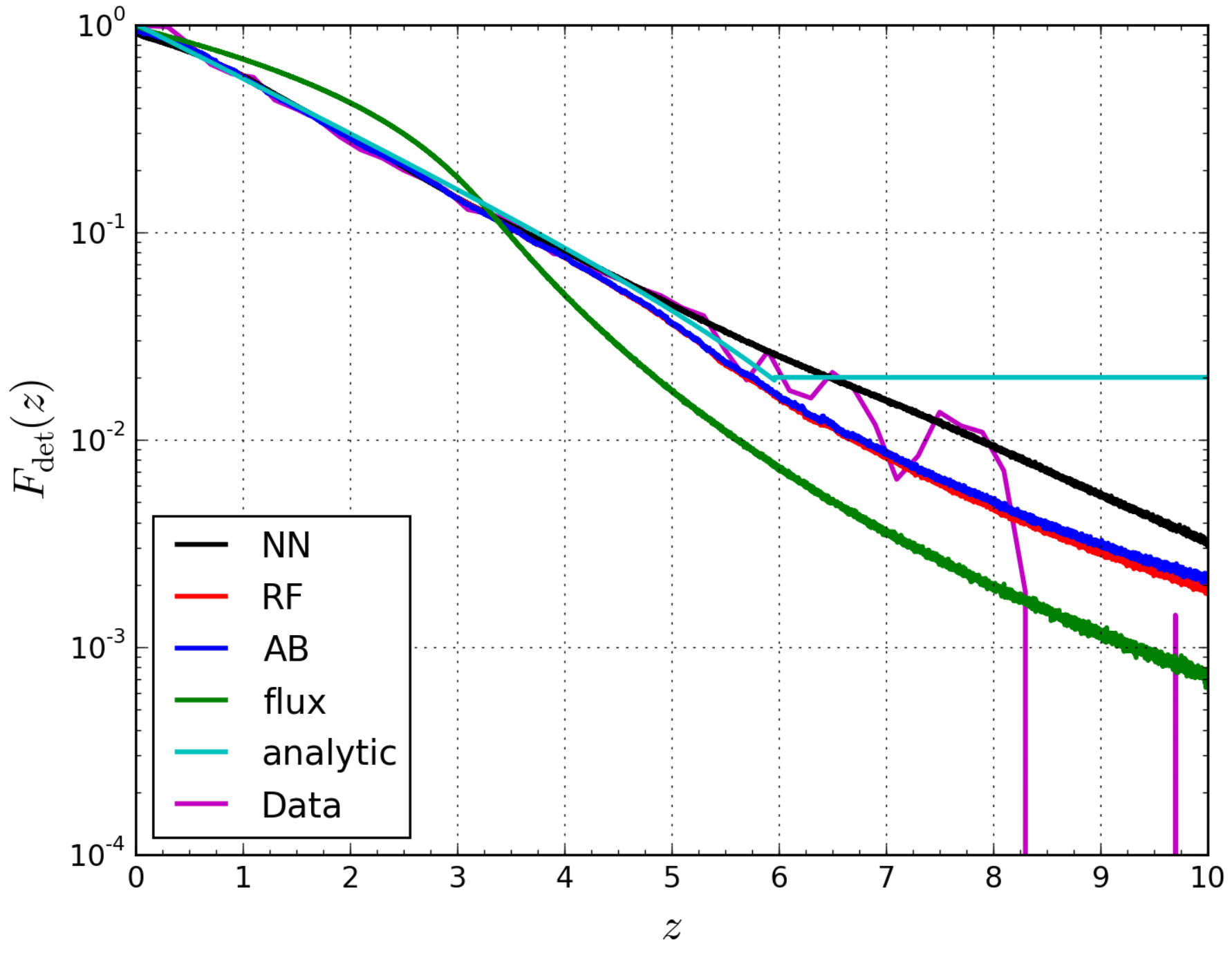}
\caption{$F_{\rm det}(z)$ as computed by the three different MLAs used as well as the constant flux cut and an analytic form used in~\cite{Howell2014}. The detection fraction of all data provided for training and validation is also shown. This is calculated under the assumption of the particular luminosity function used in this study and may change significantly for other choices of the luminosity function parameters.}
\label{fig:detection_fraction}
\end{figure}

The detection fraction (also known as detection efficiency) $F_{\rm det}(z)$ is computed in advance of the analysis by utilizing the ML models trained to reproduce the \Sw{} detection pipeline. $10^6$ GRBs are simulated at each of $10\rm{,}001$ redshift points in $[0,10]$ in order to precisely measure the average detection fraction. These points are used as the basis for a spline interpolation to compute $F_{\rm det}(z)$ at any $z$. The detection fraction as a function of $z$ from each the three models used is shown in Figure~\ref{fig:detection_fraction}. It is important to note that this $F_{\rm det}(z)$ is calculated under the assumption of the particular luminosity function used in this study; it may change significantly for other choices of the luminosity function parameters.

We also show, for comparison, the detection fraction as computed by the constant flux cut and from an analytic fit used in~\cite{Howell2014}. This was computed using the data from~\cite{Lien2014}, so we are not surprised that it matches well in the low-redshift range where there is better sampling. The flux cut has discrepancies across the entire redshift range while the analytic fit is close until $z=5.96$, after which the authors used a constant value.

These can all be compared against the detection fraction of the entire data set (training and validation) provided from using the original \Sw{} pipeline of~\cite{Lien2014}. There is less resolution and large uncertainty on this curve as there are much fewer samples ($\mathcal{O}(10^5)$ vs $\mathcal{O}(10^9)$), but we can see that RF, AB, and NN track it well.

\subsection{Model, Parameters, and Prior}\label{sec:bayesmodel}
%We employ the population generation code developed in~\cite{Lien2014} in order to simulate source GRB populations with a given set of parameters. In our initial analysis, we fix the luminosity distribution of GRBs (given by Equation~\ref{eq:luminosityfctn}) with $x=-0.65$, $y=-3.00$, and $L_{\star}=10^{52.05} \, \mathrm{erg}/\mathrm{s}$. We also fix the break point of the redshift distribution (given by Equation~\ref{eq:redshiftfctn}) with $z_1=3.60$. Thus, we leave $n_0$, $n_1$, and $n_2$ as free parameters in the following ranges:
%\begin{align}
%n_0 &\in [0.25,2.00] \, \mathrm{Gpc}^{-3}\mathrm{yr}^{-1} \nonumber \\
%n_1 &\in [1.60,4.00] \nonumber \\
%n_2 &\in [-4.00,0.00] \nonumber
%\label{eq:priors}
%\end{align}
%The prior on $n_0$ is log-uniform and the priors on $n_1$ and $n_2$ are uniform. Using Bayesian inference, we will find the best fit to the observed population of GRBs in these parameters and characterize the uncertainty in the fit with the posterior probability distribution.

In our analysis, as the detection fraction is averaged over the luminosity distribution, we hold those parameters constant with $x=-0.65$, $y=-3.00$, and $L_{\star}=10^{52.05} \, \mathrm{erg}/\mathrm{s}$. The parameters describing the redshift distribution are allowed to vary with ranges and prior distribution given in Table~\ref{tab:PEpriors}.
\begin{table}
\centering
\begin{tabular}{c | ccc}
\textbf{Parameter} & \textbf{Min} & \textbf{Max} & \textbf{Prior} \\
\hline
\hline
$n_0$ & 0.01 & 2.00 & logarithmic \\
$n_1$ & 0.00 & 4.00 & flat \\
$n_2$ & -6.00 & 0.00 & flat \\
$z_1$ & 0.00 & 10.00 & flat
\end{tabular}
\caption{Prior ranges and distributions for the redshift distribution model parameters.}
\label{tab:PEpriors}
\end{table}

The population generation code developed in~\cite{Lien2014} was used to generate simulated data for testing purposes. In addition to the above-specified parameters, we also return the total number of GRBs, $N_{\rm exp}$.

\subsection{Parameter Estimation Tests}\label{sec:bayesPE}
The BAMBI algorithm~\citep{Feroz:2008,Feroz:2009,Graff:2012} is a general-purpose implementation of the nested sampling algorithm for Bayesian inference. We use it to perform Bayesian parameter estimation, measuring the full posterior probability distribution of the model parameters.

In the ideal case, any $X\%$ credible interval calculated from the posterior distribution should contain the true parameters $\sim X\%$ of the time. We sampled a large number of parameter values from the prior and obtained a posterior distribution from simulated data generated with each. For each parameter, we then computed the cumulative fraction of times the true value was found at a credible interval of $p$ - as integrated up from the minimum value - as a function of $p$. This result was compared to a perfect one-to-one relation using the Kolmogorov-Smirnov test. All parameters passed this test, thus confirming the validity of returned credible intervals.

\begin{figure}
\centering
\includegraphics[width=0.95\columnwidth]{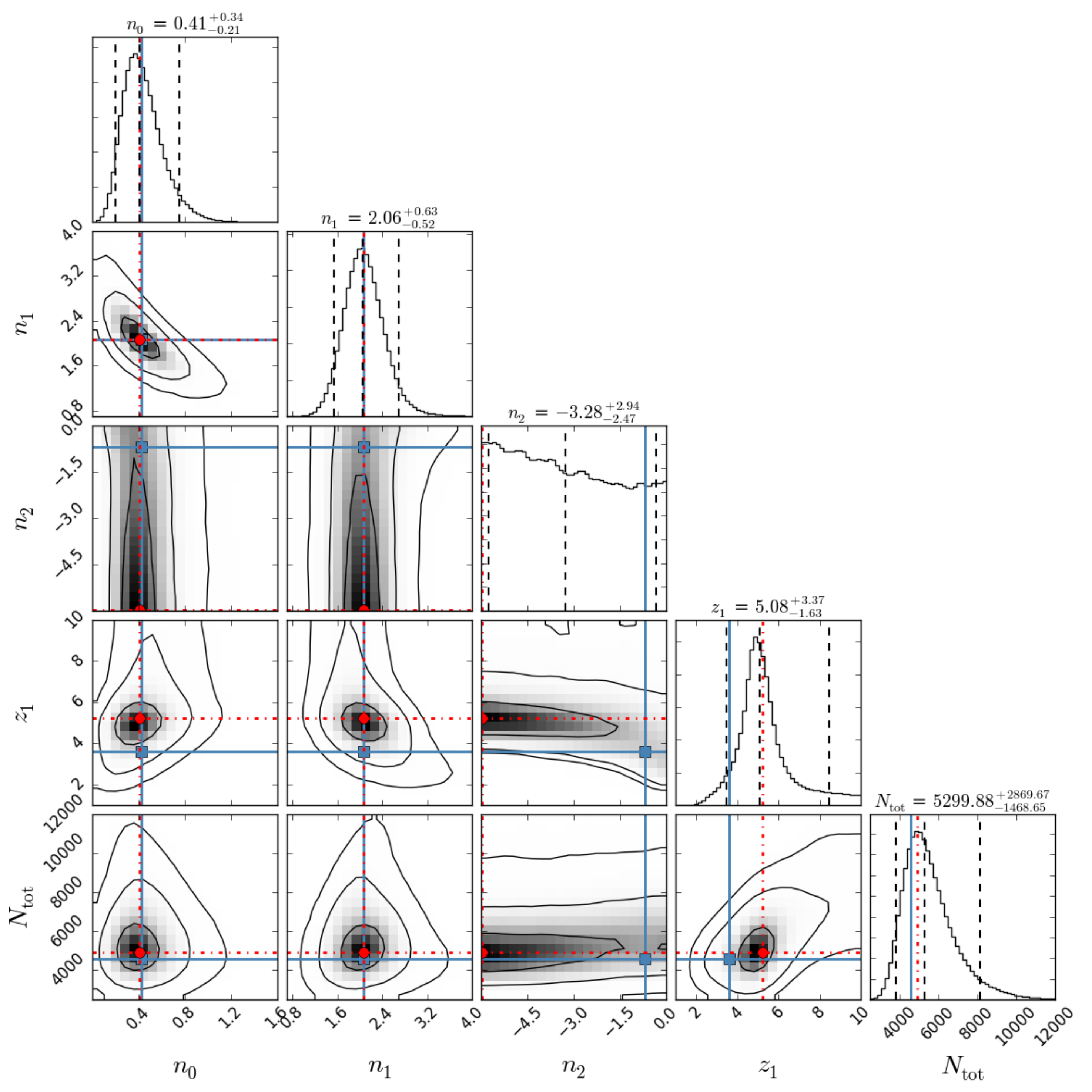}
\caption{Posterior distribution for simulated data with $\{n_0,n_1,n_2,z_1\}=\{0.42,2.07,-0.70,3.60\}$ using the random forest classifier for data generation and detection fraction. $N_{\rm tot}$ is the total number of GRBs in the Universe per year. Blue lines indicate true values and dot-dash red lines indicate maximum likelihood (i.e. best-fit) values. 2D plots show contour lines every $\sigma$ ($68\%$, $95\%$, $99\%$). Vertical dashed lines in 1D plots show $5\%$, $50\%$, and $95\%$ quantiles, with values given in the titles.}
\label{fig:bestfit_posterior_RF}
\end{figure}

\begin{figure}
\centering
\includegraphics[width=0.95\columnwidth]{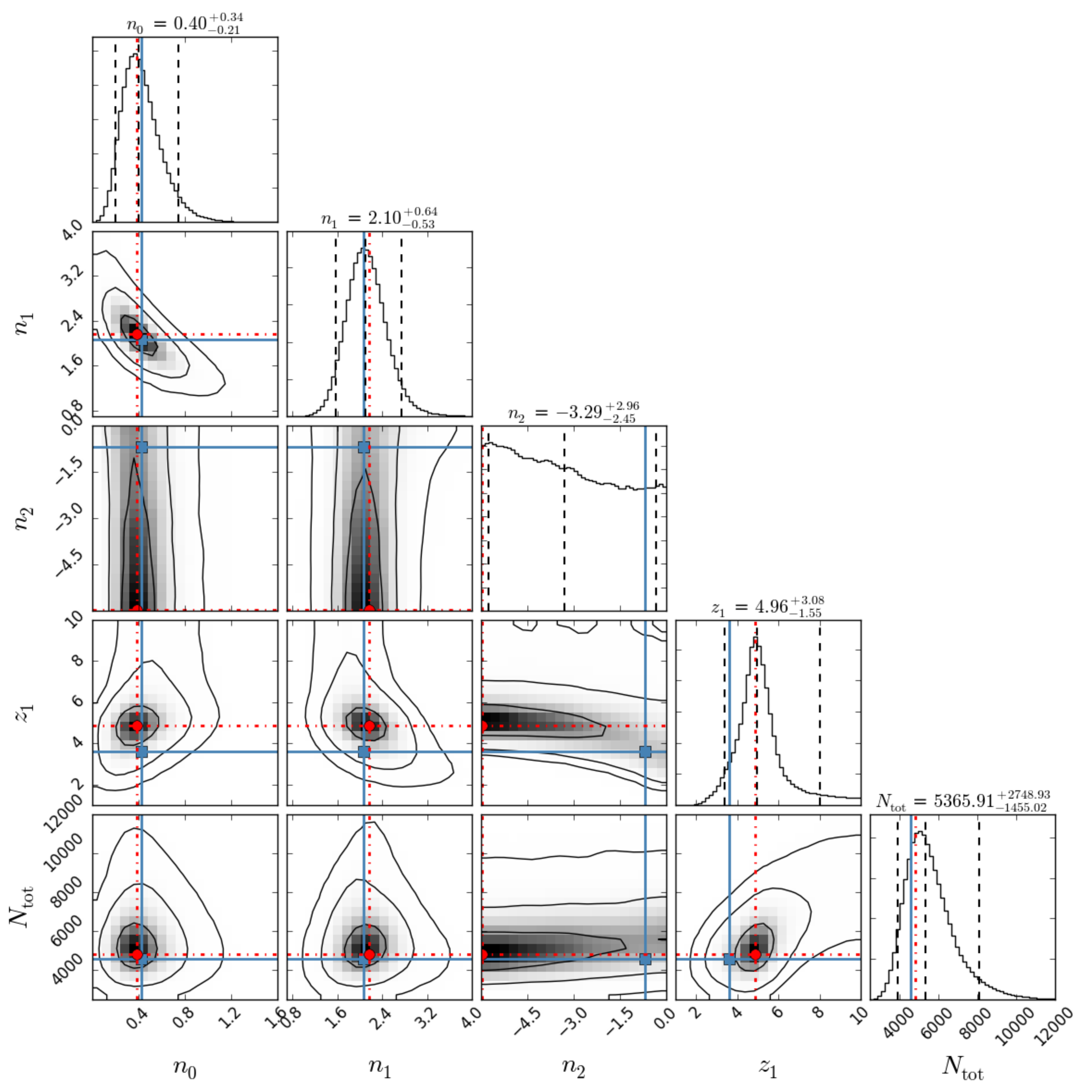}
\caption{Posterior distribution for simulated data with $\{n_0,n_1,n_2,z_1\}=\{0.42,2.07,-0.70,3.60\}$ using the AdaBoost classifier for data generation and detection fraction. Same features as Figure~\ref{fig:bestfit_posterior_RF}.}
\label{fig:bestfit_posterior_AB}
\end{figure}

\begin{figure}
\centering
\includegraphics[width=0.95\columnwidth]{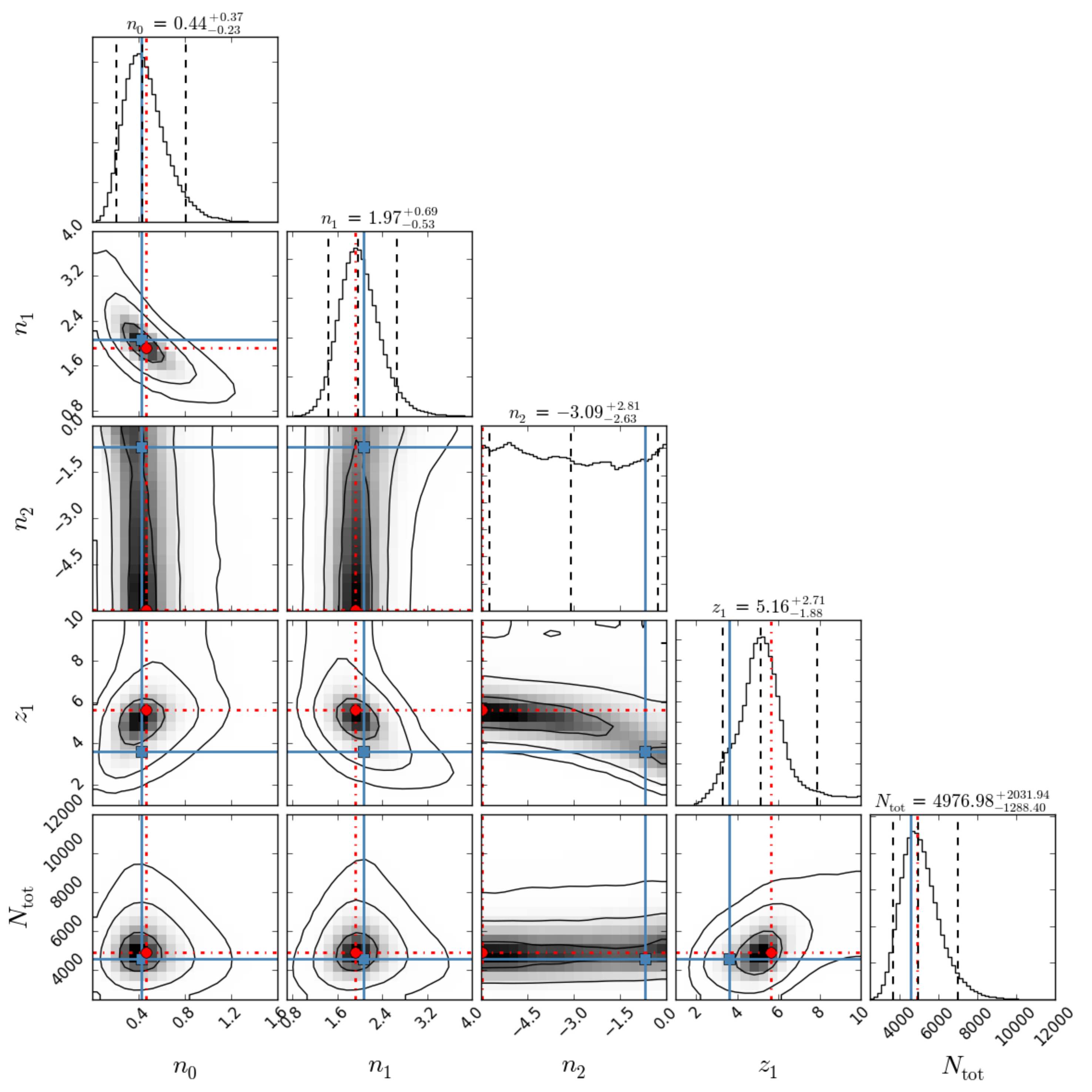}
\caption{Posterior distribution for simulated data with $\{n_0,n_1,n_2,z_1\}=\{0.42,2.07,-0.70,3.60\}$ using the \SN{} NN classifier for data generation and detection fraction. Same features as Figure~\ref{fig:bestfit_posterior_RF}.}
\label{fig:bestfit_posterior_NN}
\end{figure}

The posterior distribution for a particular realization of an observed GRB redshift distribution generated using $\{n_0,n_1,n_2,z_1\}=\{0.42,2.07,-0.70,3.60\}$ (best-fit values from~\cite{Lien2014}) is shown in Figures~\ref{fig:bestfit_posterior_RF},~\ref{fig:bestfit_posterior_AB}, and~\ref{fig:bestfit_posterior_NN} for the random forest, AdaBoost, and \SN{} NN models, respectively. While the random forest and AdaBoost posteriors are nearly identical, the \SN{} posterior has small differences due to the difference in detection fraction. However, these differences are not major. We can see that $n_2$ is effectively unconstrained due to the low number of observed GRBs with redshift greater than $z_1$. The true values are marked by the blue lines.

\begin{figure}
\centering
\includegraphics[width=0.9\columnwidth]{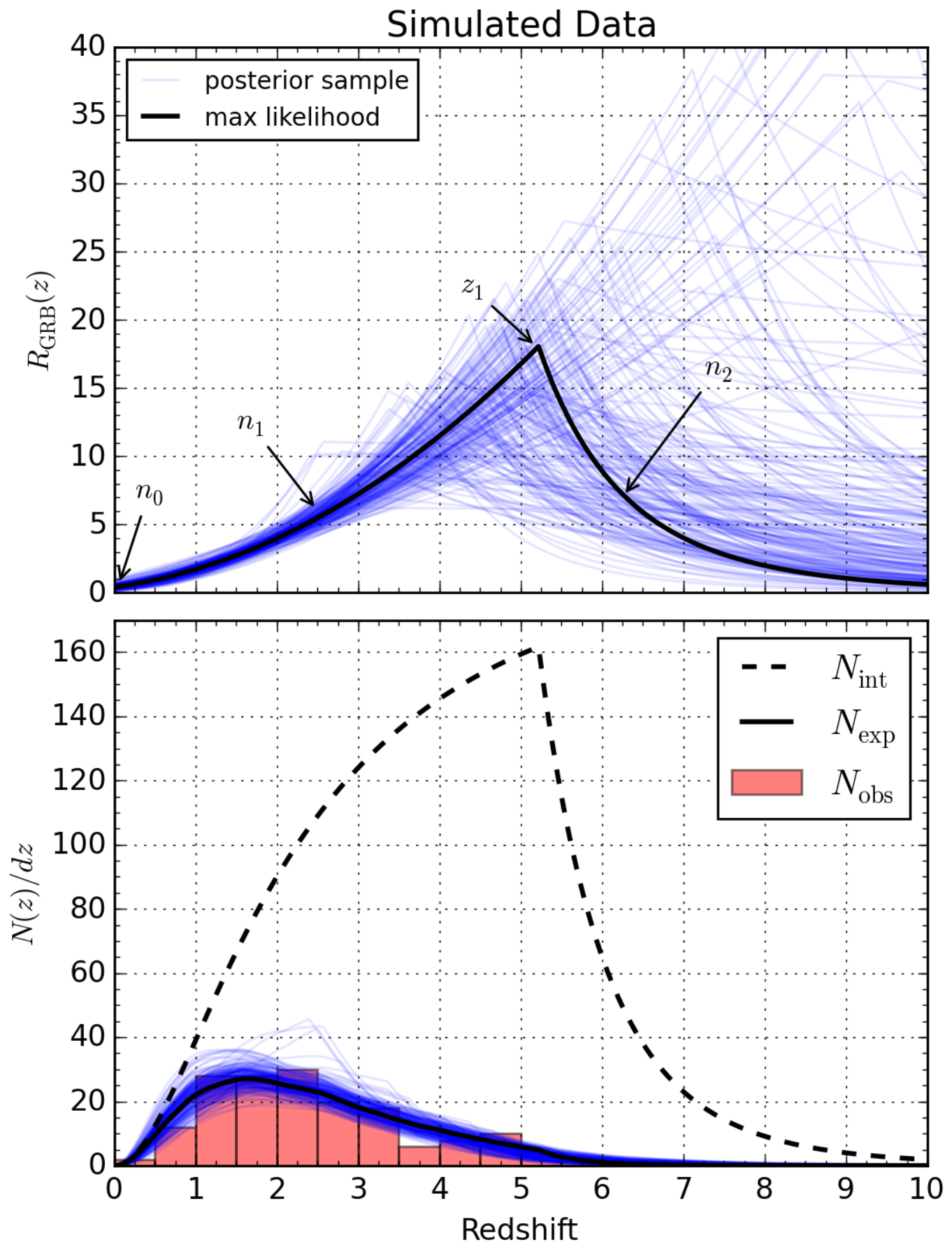}
\caption{The distribution of model predictions from the posterior (RF) for a simulated population of GRBs. $200$ models with parameters chosen randomly from the posterior are shown in light blue lines in both panels. The maximum $\mathcal{L}(\vec{n})$ point is shown in black. The upper panel shows $R_{\rm GRB}(z)$ (Equation~\eqref{eq:redshiftfctn}) and the lower panel shows $N_{\rm exp}(z)/dz$ (Equation~\eqref{eq:expected_bin_GRBs}). The lower panel also shows the simulated population of measured redshifts for observed GRBs and $N_{\rm int}(z)/dz$ for the maximum $\mathcal{L}(\vec{n})$ point in dashed black.}
\label{fig:post_model_dist_bestfit}
\end{figure}

We also plot in Figure~\ref{fig:post_model_dist_bestfit} the distribution of model predictions as specified by the posterior (from RF). In both panels, we select $200$ random models selected from the set of posterior samples (light blue lines) as well as the maximum $\mathcal{L}(\vec{n})$ point (black line). The upper panel shows $R_{\rm GRB}(z)$ (Equation~\eqref{eq:redshiftfctn}); the lower panel shows $N_{\rm exp}(z)/dz$ (Equation~\eqref{eq:expected_bin_GRBs}) and $N_{\rm int}(z)/dz$. The lower panel also plots a histogram of the simulated population of measured redshifts for observed GRBs. The upper panel clearly shows us the allowed variability in the high redshift predictions of the model; in the lower panel, we see that the detection fraction and other factors constrain this variability to consistently low predictions.

These tests show that we can trust the results of an analysis -- under the model assumptions, we can recover the true parameters of a simulated GRB redshift distribution.

\subsection{Analysis of \Sw{} GRBs}\label{sec:bayesPEreal}
In~\cite{Lien2014}, the authors use a sample of $66$ GRBs observed by \Sw{} whose redshift has been measured from afterglows only or afterglows and host galaxy observations. These observations are taken from the larger set of~\cite{Fynbo2009} and the selection is done in order to remove bias towards lower-redshift GRBs in the fraction with measured redshifts (see Section 4.1 of~\cite{Lien2014}). In our final analysis, we use these $66$ GRB redshift measurements as data that we fit with the models described in this paper.

\begin{figure}
\centering
\includegraphics[width=0.95\columnwidth]{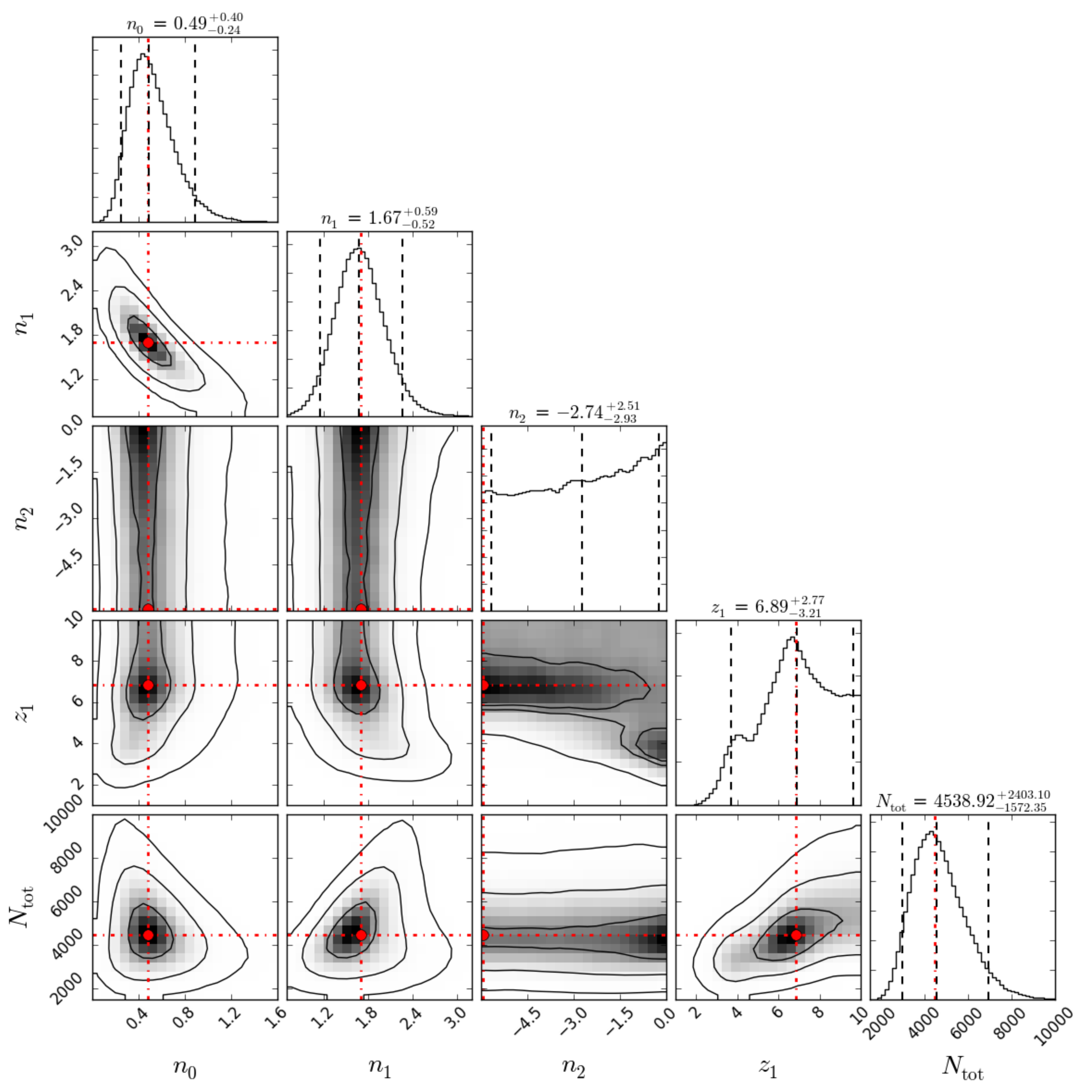}
\caption{Posterior distribution for the real set of $66$ \Sw{} GRBs using the random forest classifier for the detection fraction. $N_{\rm tot}$ is the total number of GRBs in the Universe per year. The dot-dash red lines indicate maximum likelihood (i.e. best-fit) values. 2D plots show contour lines every $\sigma$ ($68\%$, $95\%$, $99\%$). Vertical dashed lines in 1D plots show $5\%$, $50\%$, and $95\%$ quantiles, with values given in the titles.}
\label{fig:realdata_posterior_RF}
\end{figure}

\begin{figure}
\centering
\includegraphics[width=0.95\columnwidth]{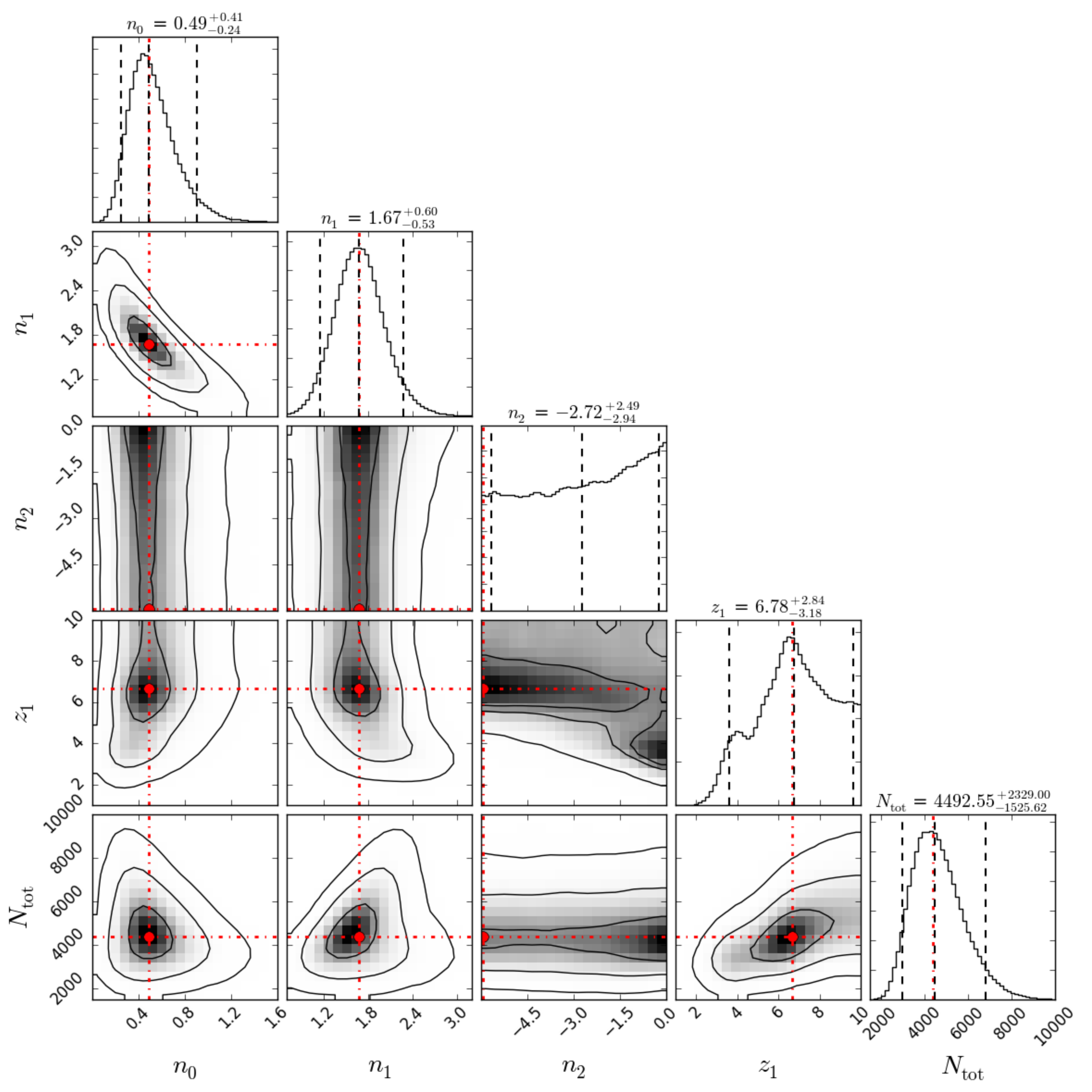}
\caption{Posterior distribution for the real set of $66$ \Sw{} GRBs using the AdaBoost classifier for the detection fraction. Similar to Figure~\ref{fig:realdata_posterior_RF}.}
\label{fig:realdata_posterior_AB}
\end{figure}

\begin{figure}
\centering
\includegraphics[width=0.95\columnwidth]{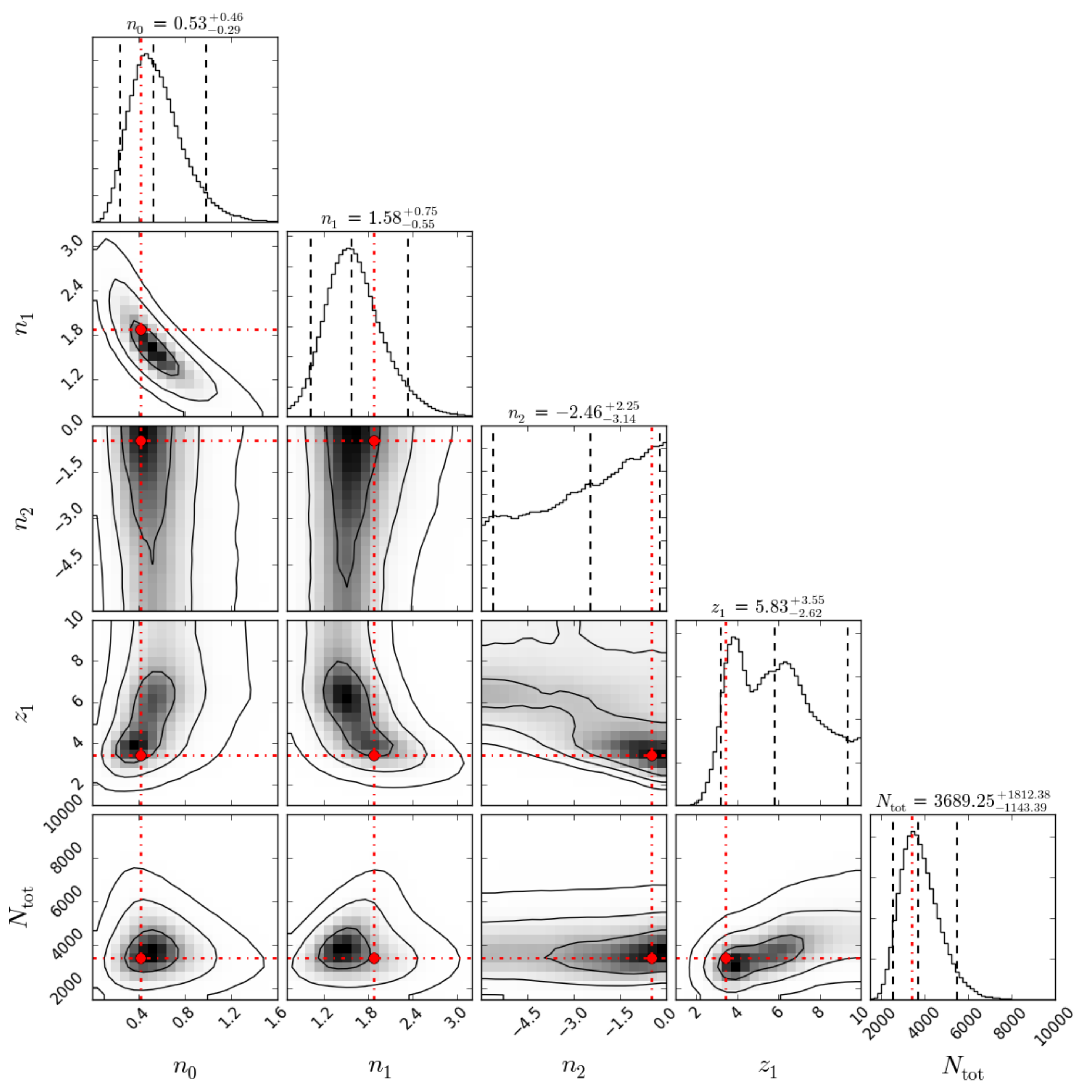}
\caption{Posterior distribution for the real set of $66$ \Sw{} GRBs using the \SN{} NN classifier for the detection fraction. Similar to Figure~\ref{fig:realdata_posterior_RF}.}
\label{fig:realdata_posterior_NN}
\end{figure}

\begin{figure}
\centering
\includegraphics[width=0.9\columnwidth]{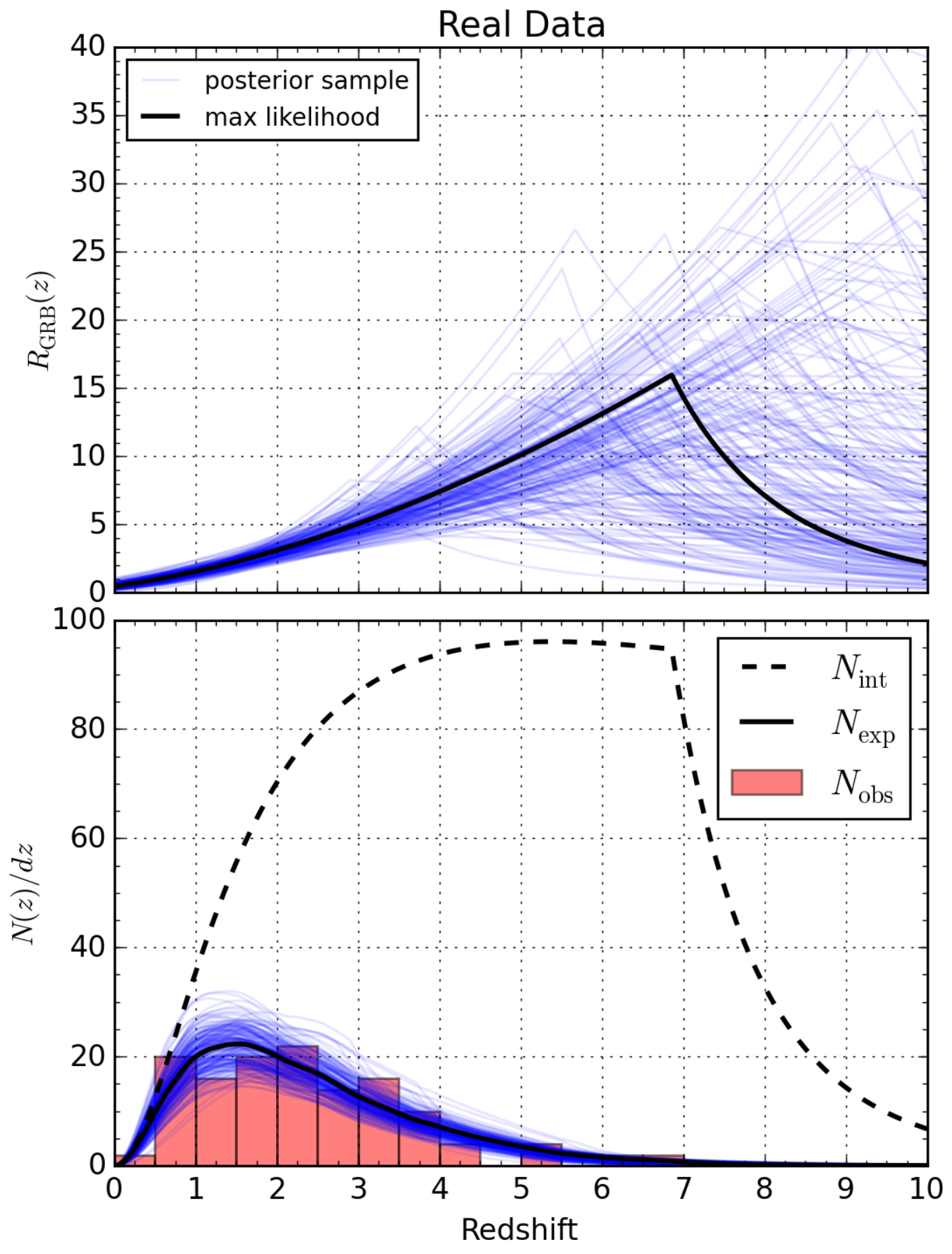}
\caption{The distribution of model predictions from the posterior (RF) for the real set of $66$ \Sw{} GRBs~\citep{Fynbo2009}. $200$ models with parameters chosen randomly from the posterior are shown in light blue lines in both panels. The maximum $\mathcal{L}(\vec{n})$ point is shown in black. The upper panel shows $R_{\rm GRB}(z)$ (Equation~\eqref{eq:redshiftfctn}) and the lower panel shows $N_{\rm exp}(z)/dz$ (Equation~\eqref{eq:expected_bin_GRBs}). The lower panel also shows the distribution of measured redshifts for observed GRBs and $N_{\rm int}(z)/dz$ for the maximum $\mathcal{L}(\vec{n})$ point in dashed black.
%The distribution of model predictions from the posterior as in Figure~\ref{fig:post_model_dist_bestfit} for the real set of $66$ \Sw{} GRBs using the random forest classifier for the detection fraction.
}
\label{fig:post_model_dist_realdata}
\end{figure}

Using random forests, AdaBoost, and neural network ML models for the detection fraction, we find posterior probability distributions for $n_0$, $n_1$, $n_2$, and $z_1$, as seen in Figures~\ref{fig:realdata_posterior_RF},~\ref{fig:realdata_posterior_AB}, and~\ref{fig:realdata_posterior_NN}, respectively. The maximum likelihood estimate and posterior probability central $90\%$ credible interval are given in Table~\ref{tab:realdata_max_RF}. We also plot in Figure~\ref{fig:post_model_dist_realdata} the distribution of model predictions as specified by the posterior (from RF) as we did in Figure~\ref{fig:post_model_dist_bestfit} for the test population.

Parameters $n_0$, $n_1$, and $N_{\rm tot}$ show mostly Gaussian marginal distributions and some correlation between $n_0$ and $n_1$ -- larger values of the former lead to lower values of the latter in order to maintain a constant value for $N_{\rm tot}$ and similar values at the peak of the observed distribution. The data do not strongly constrain the high redshift part of the distribution, namely the $n_2$ parameter. The upper panel of Figure~\ref{fig:post_model_dist_realdata} clearly shows us the allowed variability in the high redshift predictions of the model; in the lower panel, we see that the detection fraction and other factors constrain this variability to consistently low predicted numbers of GRB observations. We see a double-peak in $z_1$, not the clear single peak seen in the simulated data. One peak occurs around $z_1 \approx 3.6$, the best-fit value from~\cite{Lien2014} and is more prominent when using the NN model. This shows a sensitivity to the detection fraction for this set of GRB observations. A hint of this can be seen in the posterior plots of Section~\ref{sec:bayesPE} --  Figures~\ref{fig:bestfit_posterior_RF},~\ref{fig:bestfit_posterior_AB}, and~\ref{fig:bestfit_posterior_NN}. All measured parameters are consistent with the best-fit values found by~\cite{Lien2014}.

\begin{table}
\centering
\begin{tabular}{cc | cc}
\textbf{Parameter} & \textbf{Method} & \textbf{Max Like} & $\mathbf{90\%}$ \textbf{CI} \\
\hline
\hline
 & RF & 0.480 & [0.247, 0.890] \\
$n_0$ & AB & 0.489 & [0.249, 0.902] \\
 & NN & 0.416 & [0.238, 0.986] \\
\hline
 & RF & 1.700 & [1.155, 2.261]\\
$n_1$ & AB & 1.681 & [1.146, 2.273] \\
 & NN & 1.875 & [1.030, 2.334] \\
\hline
 & RF & -5.934 & [-5.675, -0.238]\\
$n_2$ & AB & -5.950 & [-5.665, -0.230] \\
 & NN & -0.483 & [-5.598, -0.217] \\
\hline
 & RF & 6.857 & [3.682, 9.654] \\
$z_1$ & AB & 6.682 & [3.603, 9.622] \\
 & NN & 3.418 & [3.215, 9.385] \\
\hline
 & RF & 4455 & [2967, 6942] \\
$N_{\rm exp}$ & AB & 4392 & [2967, 6822] \\
 & NN & 3421 & [2546, 5502] 
\end{tabular}
\caption{Maximum likelihood (i.e. best-fit) estimates and central $90\%$ credible intervals for the redshift distribution parameters as fit to the real set of $66$ \Sw{} GRBs~\citep{Fynbo2009,Lien2014} using each of the MLAs.}
\label{tab:realdata_max_RF}
\end{table}

%From these posterior distributions we can see that $n_2$ is largely unconstrained, showing only slightly more support at larger values. This is due to the fact that the contribution at higher redshifts is suppressed due to the much lower detection fraction, as seen in Figure~\ref{fig:post_model_dist_realdata}.

\subsection{Computational Cost}\label{sec:bayesCompCost}
The main computational costs of this entire analysis procedure were:
\begin{enumerate}
\item Producing the training data
\item Performing MLA model fitting and hyper-parameter optimization
\item Using the MLA models to compute the detection fraction.
\end{enumerate}
These steps are in roughly decreasing order of cost, from CPU weeks to days. However, all three are one-time initialization costs and can be run massively parallel to reduce wall-time.

After this initialization is complete, however, subsequent analysis of real or simulated data is performed extremely quickly. A single likelihood evaluation takes $<0.1$ ms, meaning that a Bayesian analysis can be computed in less than a minute on a laptop. Providing this same kind of accurate measurement of the detection fraction without the MLAs would take orders of magnitude more time; while $\mathcal{O}(10^5)$ samples were used for training the MLA mdoels, $\mathcal{O}(10^{10})$ evaluations were used in measuring the detection fraction as a function of redshift. The precision of the detection fraction would need to be reduced significantly to make the overall cost comparable. Furthermore, we now are equipped with accurate models of the \Sw{} detection algorithm.

%++++++++++++++++++++++++++++++++++++++++++++++++++++++++++++++++++++++
%
%  COMPARISON TO PREVIOUS WORK
%
%++++++++++++++++++++++++++++++++++++++++++++++++++++++++++++++++++++++

\section{Comparison to Previous Work}\label{sec:comparison}

We have developed a machine learning algorithm (simulator) for the detailed \Sw{} BAT long GRB pipeline simulator developed in~\cite{Lien2014}. These techniques allow us to complete a thorough Bayesian analysis of the long GRB rate redshift dependence using the~\cite{Fynbo2009} data set, improving on the more coarsely sampled study in~\cite{Lien2014}. Our results are compatible with those from~\cite{Lien2014} with tight agreement for lower redshifts up to $z\sim 4$ with compatible results and relatively narrow distributions for our $n_1$ and $n_0$ rate parameters. We find values of $n_0 \sim 0.48^{+0.41}_{-0.23} \ {\rm Gpc}^{-3} {\rm yr}^{-1}$ and $n_1 \sim 1.7^{+0.6}_{-0.5}$, consistent with the best-fit values of $n_0 = 0.42$ and $n_1 = 2.07$ from~\cite{Lien2014}. For larger redshifts the model is less constrained; $n_2$ spans the prior range and $z_1$ is significantly constrained only at the low-$z$ end. Our general agreement with~\cite{Lien2014} supports their identification of differences between the long GRB redshift distribution and estimates of the star formation rate~\citep{Hopkins2006}. Though our analysis indicates that the~\cite{Fynbo2009} data do not provide strong constraints on the rate at high redshift the results seem to indicate significant differences for $z<4$. A follow-up Bayesian analysis comparing with a two-break model would allow a more direct comparison with SFR models. We can also note how our results compare with several other studies which use GRB observations and subsequent redshift measurements in order to estimate the redshift or luminosity distribution of GRBs in the Universe. 
%In this section we will briefly review a selection of similar work that has been performed. These studies used GRB observations and subsequent redshift measurements in order to estimate the redshift or luminosity distribution of GRBs in the Universe. These are in addition to the obvious comparison between our work and that of~\cite{Lien2014}. Studies reviewed are those of~\cite{Butler10},~\cite{Wanderman10},~\cite{Salvaterra12},~\cite{Howell2014},~\cite{Yu15}, and~\cite{Petrosian15}.

The paper by~\cite{Butler10} used an extensive set of GRBs both with and without redshift measurements to fit intrinsic distributions for GRB redshift, luminosity, peak flux, and more. This fitting was performed using PyMC, a python package for Markov chain Monte Carlo analyses, marginalizing over all redshifts when no measurement is available; the log-likelihood function used is un-binned, similar to the one used in our study. The detection fraction (a.k.a. detection efficiency) used by~\cite{Butler10}, however, is a probability dependent solely on the photon count rate. Their results for $n_1$, $n_2$, and $z_1$ are consistent with $90\%$ confidence intervals that we measure.

%The work by~\cite{Wanderman10} fits the redshift and luminosity distributions of GRBs using \Sw{} GRB observations. They use maximum likelihood estimation (MLE) with a log-likelihood binned in redshift and luminosity. The detection efficiency is modeled as a probability derived solely from the peak flux of the GRB and the authors take into account the probability of a GRB's redshift being measured following detection, also a function of the peak flux. \note{Results consistent?}

\cite{Wanderman10} performs a careful study of the GRB rate and luminosity distribution via a Monte-Carlo approach. This study adopts an empirical probability function to determine whether a burst is detectable based on the peak flux. In addition, they also introduce an empirical function to estimate the probability of obtaining a redshift measurement based on the GRB peak flux. Since we adopt the same functional form as~\cite{Wanderman10}, it is possible to compare the values of the same parameters. However, in this paper we quantify the parameter uncertainties of the GRB rate, and assume an un-changed luminosity function from~\cite{Lien2014}, which is different the one found in~\cite{Wanderman10}. The parameters found by~\cite{Wanderman10} are $n0 \sim 1.25$, $n1 \sim 2.07$, and $n_2 \sim -1.36$ (as listed in Table 2 of by~\cite{Wanderman10}). These values of $n_1$ and $n_2$ are consistent with our findings; the value of $n_0$ is at the upper end of our range, but this difference is likely due to the difference in luminosity distribution.

%\cite{Salvaterra12} similarly performs a MLE analysis, but on a different sample of GRBs. They fit only the intrinsic luminosity distribution of GRBs, using a redshift distribution proportional to the star formation rate (SFR). The likelihood function used is the same used in our work but~\cite{Salvaterra12} use both redshift and flux information, marginalizing over a flat distribution in $z$ if a value is not measured for a particular GRB. The detection efficiency of \Sw{} is modeled as a threshold on the GRB flux. \note{Results consistent?}

\cite{Salvaterra12} constructs a sub-sample of \Sw{} long GRBs that is complete in redshift by selecting bursts that satisfy certain observational criteria that are optimal for follow-up observations. In addition, these authors select only bright bursts with $1$-s peak photon fluxes greater than $2.6$ photons $\rm s^{-1} \ \rm cm^{-2}$, in order to achieve a high completeness of $90\%$ in redshift measurements. They use this sub-sample to estimate the luminosity function and GRB rate via maximum likelihood estimation -- using the same likelihood as our study and marginalizing over a flat $z$ distribution if no value was measured for a GRB --, and found that either the rate or the luminosity function is required to evolve strongly with redshift, in order to explain the observational data. The \Sw{} detection efficiency is modeled as a threshold on the GRB flux. The rate model fits of~\cite{Salvaterra12} are not directly comparable to ours due to a different functional form based off of the SFR.

The study of~\cite{Howell2014} takes advantage of some of the work done by~\cite{Lien2014} in using the detection efficiency computed from simulated GRB populations. The authors perform a time-dependent analysis that considers the rarest events -- the largest redshift or the highest peak flux -- and how these values progress over observational time. These are used to fit the intrinsic redshift and luminosity distributions of GRBs and infer $90\%$ confidence intervals.~\cite{Howell2014} measures a local GRB rate density consistent with our constraints on $n_0$. Other rate parameters were held fixed to values obtained by~\cite{Lien2014} and are thus also consistent with our measurements.

%Lastly, the research reported in~\cite{Yu15} fits the intrinsic redshift and luminosity distributions using Lynden-Bell's $c^-$ method. They consider a redshift dependency on the luminosity distribution and compare the GRB redshift rate to the SFR. A flux threshold is used to model the \Sw{} detection efficiency. \note{Results consistent?}

\cite{Yu15} and~\cite{Petrosian15} use sub-sets of observed GRBs at different redshifts to construct a more complete GRB sample and account for observational biases. This method is called Lynden-Bell’s $\rm c^{-}$ method. Each sub-sample is selected based on the minimum detectable GRB luminosity at each redshift. Both of these studies find significant luminosity evolution and a rather high GRB rate at low redshift in comparison to the one expected from previous star-formation rate measurements. However, as noted in~\cite{Yu15} and~\cite{Petrosian15}, several additional selection effects can be the cause of this discrepancy, including the potential bias toward redshift measurements of nearby GRBs and those with bright X-ray and optical afterglows. The rate evolution found by~\cite{Yu15} is not consistent with our results at low redshift, but is consistent at high redshift due to the large uncertainty in measuring $n_2$.

Our study is able to improve upon the methodology of these studies and may be extended to cover the same breadth of GRB source models to be fit. These improvements are not the same for all, but in summary involve using a fully Bayesian model fitting procedure with a likelihood function that does not involve any binning of observations. Furthermore, the detection efficiency of the \Sw{} BAT detector can be better modeled using ML techniques that incorporate all available information (marginalizing over parameters not under consideration) than with probabilities dependent solely on the flux or photon counts. With both of these, not only will we be able to extract as much information as possible out of GRB detections and follow-up observations, but such analyses will incur minimal modeling bias while maintaining computational speed.

%++++++++++++++++++++++++++++++++++++++++++++++++++++++++++++++++++++++
%
%  CONCLUSION
%
%++++++++++++++++++++++++++++++++++++++++++++++++++++++++++++++++++++++

\section{Summary and Conclusions}\label{sec:conclusion}

%\subsection{Summary of Results and Conclusions}\label{sec:conclusion2}
We have built a set of models emulating the \Sw{} BAT detection algorithm for long GRBs using machine learning. Using a large set of simulated GRBs from the work of~\cite{Lien2014} as training data, we used the random forest, AdaBoost, support vector machine, and neural network algorithms to optimize, fit, and validate models that simulate the \Sw{} triggering algorithm to high accuracy. RF and AdaBoost perform best, achieving accuracies of $97.5\%$; NNs and SVMs have accuracies of $96.9\%$ and $94.7\%$, respectively. These all out-perform a threshold in GRB flux, which has an accuracy of $89.6\%$. The improved faithfulness to the full \Sw{} triggering removes potential sources of bias when performing analyses based on the model.

Using these models, we computed the detection fraction (efficiency) of \Sw{} as a function of redshift for a fixed luminosity distribution. Using this empirical detection fraction and a model for the GRB rate given by Equation~\eqref{eq:redshiftfctn}, we fit the model parameters on both simulated redshift measurements and on the redshifts reported by~\cite{Fynbo2009}. We find best-fitting values and $90\%$ credible intervals as reported in Table~\ref{tab:realdata_max_RF} for each of the top three MLAs. These, expectedly, are consistent with values found by~\cite{Lien2014}.

After incurring the initial costs of generating training data, fitting the models, and computing the detection fraction, we are able to perform Bayesian parameter estimation extremely rapidly. This allows us to explore the full parameter space of the model and determine not only the best-fit parameters but also the uncertainty and degeneracies present.

\section{Future Work}\label{sec:future}
In performing this analysis, we identified several potential avenues for further work to improve our model and extend the analysis performed. Hence, we see this work as just the first step in advancing GRB research. The easiest extension is to analyze different samples of measured GRB redshifts that may contain different selection biases.

To improve our ML models, we would like to continue building on our training data set and making it more agnostic with respect to GRB parameters. This would allow for better modeling of the \Sw{} detection algorithm and its dependency on different GRB characteristics. The GRB rate model can also be expanded to include a second break point in redshift. This would allow for more direct comparison with most fits of the SFR that use a double-broken power-law model. Bayesian model selection could then be used to compare these and other models.

The analyses can be extended to fitting the intrinsic luminosity distribution by including GRB luminosity or flux in the detection fraction, $F_{\rm{det}}\left(\log_{10}(L),z\right)$ or $F_{\rm{det}}\left(\Phi\left(\log_{10}(L),z\right),z\right)$. The likelihood function can then jointly describe both the luminosity and redshift distributions -- including luminosity distribution evolution with redshift -- by analyzing measured GRB fluxes and redshifts; the redshift distribution can be marginalized over if there is no measured value for a particular GRB. The likelihood function can also be modified to account for known selection biases, including the probability of measuring a redshift for each GRB.

Beyond improving and extending the model used in this paper, a similar analysis can be performed for the study of short GRBs detected by \Sw{} and other detectors. This work has demonstrated the value of machine learning for GRB data analysis and the algorithms and techniques may be extended to other problems in GRB follow-up and analysis.

\section*{Acknowledgements}

The authors would like to thank Brad Cenko, Judith Racusin, and Neil Gehrels for helpful discussions.
PG acknowledges support from NASA Grant NNX12AN10G and an appointment to the NASA Postdoctoral Program at the Goddard Space Flight Center, administered by Oak Ridge Associated Universities through a contract with NASA.
%\note{AL acknowledgements here.}
JB acknowledges support from NASA Grant ATP11-00046.

\bibliographystyle{plainnat}
\bibliography{swiftML}

\end{document}